\documentclass[aip,jmp,reprint,onecolumn]{revtex4-1}

\draft 

\usepackage{graphicx}
\usepackage{amsmath}
\usepackage{amssymb}
\usepackage{bm}
\numberwithin{equation}{section}

\renewcommand\theequation{\arabic{section}.\arabic{equation}}
\allowdisplaybreaks

\begin{document}


\title{Formulation of a unified method for low- and high-energy expansions in the analysis of reflection coefficients for~one-dimensional Schr\"odinger equation
\vspace{0.2cm}}

\author{Toru Miyazawa}

\email{toru.miyazawa@gakushuin.ac.jp}
\maketitle
\vspace{-0.8cm} \qquad \ \ 
{\itshape \small Department of Physics, Gakushuin University, Tokyo 171-8588, Japan}

\vspace{0.4cm}
\leftskip=1.3cm 
\rightskip=0.8cm
\noindent
We study low-energy expansion and high-energy expansion of reflection coefficients  for one-dimensional Schr\"odinger equation, from which expansions of the Green function can be obtained. Making use of the equivalent Fokker-Planck equation, we develop a generalized formulation of a method for deriving these expansions in a unified manner. In this formalism, the underlying algebraic structure of the problem can be clearly understood, and the basic formulas necessary for the expansions can be derived in a natural way. 
We also examine the validity of the expansions 
for various asymptotic behaviors of the potential at spatial infinity.

\leftskip=0cm
\rightskip=0cm

\vspace{0.5cm}

\section{Introduction}
Our object of study in this paper is the one-dimensional steady-state Schr\"odinger equation 
\begin{equation}
-\frac{d^2}{d x^2} \psi(x) + V_\mathrm{S}(x) \psi(x)=k^2 \psi(x),
\end{equation}
where $V_{\rm S}$ is a real-valued function, and $k$ is a complex number in the closed upper half plane ($\mathrm{Im}\,k \geq 0$).
We assume that the origin of the energy scale ($k^2=0$) is set at the energy of the ground state.
[When dealing with a Schr\"odinger operator $-d^2/dx^2 + U_\mathrm{S}(x)$ which has a ground state with energy $E_0 \neq 0$, the ground-state energy can be shifted to zero by defining $V_\mathrm{S}(x) \equiv U_\mathrm{S}(x) - E_0$.]
In other words, $k^2$ corresponds to the energy relative to the ground state. 
The ground state may either be a bound state or a continuum state.

It is well known that (1.1) is equivalent to the Fokker-Planck equation\cite{risken}
\begin{equation}
-\frac{d^2}{d x^2} \phi(x) -  \frac{d}{d x}\left\{\left[\frac{d}{dx} V(x)\right] \phi(x)\right\}=k^2 \phi(x),
\end{equation}
which describes diffusion in a potential $V(x)$. We define
\begin{equation}
f(x) \equiv -\frac{1}{2} \frac{d}{d x} V(x).
\end{equation}
Equations (1.1) and (1.2) are related by 
\begin{equation}
\psi(x)=e^{V(x)/2} \phi(x), \qquad V_\mathrm{S}(x)=[f(x)]^2 + \frac{d}{dx}f(x).
\end{equation}
Substituting $\phi=e^{-V/2} \psi$ into (1.2) yields Schr\"odinger equation (1.1). 
It is also easy to see that $\psi_0(x) \equiv e^{-V(x)/2}$ is the ground-state wavefunction satisfying (1.1) with $k=0$.

Let $U(x,y;k)$ be the $2 \times 2$ matrix satisfying the differential equation
\begin{equation}
\frac{\partial}{\partial x}
U(x,y;k)=
\begin{pmatrix}
-i k & f(x) \\
f(x) & i k \\
\end{pmatrix}
U(x,y;k)
\end{equation}
with the boundary condition
$U(y,y;k)=I$ (identity matrix).
Here, $f$ is the the function defined by (1.3). 
The elements of $U$ can be written with two functions $\alpha$ and $\beta$ as
\begin{equation}
U(x,y;k)\equiv
\left(
\begin{array}{cc}
\alpha(x,y;k) & \beta(x,y;-k) \\
\beta(x,y;k) & \alpha(x,y;-k) \\
\end{array}
\right).
\end{equation} 
We define the transmission coefficient $\tau$, the left reflection coefficient $R_l$, and the right reflection coefficient $R_r$ as
\begin{equation}
\tau(x,y;k)\equiv \frac{1}{\alpha(x,y;k)},
\qquad
R_l(x,y;k)\equiv -\frac{\beta(x,y;-k)}{\alpha(x,y;k)},
\qquad
R_r(x,y;k)\equiv \frac{\beta(x,y;k)}{\alpha(x,y;k)}.
\end{equation}
(The reason why they are called the transmission and reflection coefficients is explained in Appendix~A of Ref.~\onlinecite{periodic}.)
If $V(x)$ is sufficiently well behaved at infinity, we can define the reflection coefficients for semi-infinite intervals as
\begin{equation}
R_r(b, -\infty;k) \equiv \lim_{a \to -\infty} R_r(b,a;k),
\qquad
R_l(\infty, a;k) \equiv \lim_{b \to +\infty} R_l(b,a;k).
\end{equation}
When $k$ is a real number, it may happen that the limits in (1.8) do not exist. In such cases, we interpret (1.8) as the limit approached from the upper half plane,
\begin{equation}
R_r(b,-\infty;k) \equiv \lim_{\epsilon \downarrow 0} \lim_{a \to -\infty} R_r(b,a;k+ i\epsilon),
\qquad
R_l(\infty,a;k) \equiv \lim_{\epsilon \downarrow 0} \lim_{b \to +\infty} R_l(b,a;k+ i \epsilon).
\end{equation}

Let $ G_\mathrm{S}(x,y;k)$ be the Green function of Schr\"odinger equation (1.1) satisfying
\begin{equation}
\left[\frac{\partial^2}{\partial x^2}-V_\mathrm{S}(x) +k^2 \right]G_\mathrm{S}(x,y;k)
=\delta(x-y)
\end{equation}
with the boundary conditions $G_\mathrm{S}(x,y;k) \to 0$ as $\vert x - y \vert \to \infty$ for $\mathrm{Im}\,k>0$. [For $\mathrm{Im}\,k=0$, we define $G_\mathrm{S}(x,y;k) \equiv \lim_{\epsilon \downarrow 0}G_\mathrm{S}(x,y;k+i \epsilon)$.] We can express this $G_\mathrm{S}$ as\cite{expressions} 
\begin{equation}
G_\mathrm{S}(x,y;k)=
\frac{1}{2 i k\sqrt{[1-S(x,k)][1-S(y,k)]}}
\exp\left[
i k(x-y)- i k\int_y^x S(z,k)\,dz
\right],
\end{equation}
where we have assumed (without loss of generality) that $x \geq y$ and defined
\begin{equation}
S(x,k)\equiv S_r(x,k)+S_l(x,k),
\end{equation}
\begin{equation}
S_r(x,k) \equiv \frac{R_r(x,-\infty;k)}{1+R_r(x,-\infty;k)}, 
\qquad
S_l(x,k) \equiv \frac{R_l(\infty;x;k)}{1+R_l(\infty,x;k)}.
\end{equation}

Asymptotic behavior of solutions of the Schr\"odinger equation in low- and high-energy regions has been an important object of study for many years.\cite{deift, yafaef, bolle, newton, klaus1, aktosun3, rybkin1, verde, harris1, hinton1, rybkin2, ramond, costin} Since the Green function can be expressed solely in terms of $R_r(x,-\infty;k)$ and $R_l(\infty,x;k)$ as shown in (1.11), all the information about the behavior of solutions can be obtained through the analysis of the reflection coefficients.
The study of the reflection coefficients is also important in spectral theory and inverse problems. 
In particular, $S_r(x,k)$ and $S_l(x,k)$ defined by (1.13) are closely related to the Weyl-Titchmarsh $m$-functions, which play a significant role in spectral analysis of the Schr\"odinger operator.\cite{atkinson, everitt, danielyan, bennewitz, hinton2, harris2}
 (The relation between $S_r$, $S_l$ and the $m$-functions is discussed in Appendix~C of Ref.~\onlinecite{periodic}.)  
In this paper, we investigate low- and high-energy expansions of $R_r(x,-\infty;k)$.
The results for $R_l(\infty,x;k)$ can be obtained in the same way.

To study low- and high-energy expansions, the Fokker-Planck equation is more suitable than the Schr\"odinger equation itself because the structure of the problem becomes more transparent for the Fokker-Planck equation.
By using the Fokker-Planck potential $V$,  we can carry out the analysis more systematically, and more explicit expressions can be obtained, than by using $V_\mathrm{S}$. 
In this paper, we deal with expressions in terms of $V(x)$ or $f(x)$.

We assume that $V(x)$ is a piecewise continuously differentiable\cite{note1} real-valued function.
Our method is applicable to $V(x)$ with various asymptotic behaviors at spatial infinity, including the cases where $V(x)$ is finite or $\pm \infty$ as $x \to \pm \infty$. Corresponding to such $V(x)$, the Schr\"odinger potential $V_\mathrm{S}(x)$ is finite or $+\infty$ as $x\to \pm \infty$. [We do not consider the cases where $\lim_{x \to \pm \infty} V_\mathrm{S}(x)=-\infty$.]
We will also deal with asymptotically periodic potentials. 
Specific conditions on the potential will be given when they become necessary.

In the formalism based on the Fokker-Planck equation, we can express $R_r(x,-\infty;k)$ in terms of a linear operator. The high- and low-energy expansions of the reflection coefficient then reduce to the expansions of this operator. This method has been studied in the previous papers.\cite{analysis, high, low, asymptotic}
However, the derivation of the expansions in these papers was not rigorous.
The origin of the basic formulas used in these works was not clear, and the meaning of the method has not been fully understood. 
It is the aim of the present work to construct a generalized formulation which provides a clear overall view of the problem and which gives a rigorous justification to the intuitive arguments of the previous papers. In this generalized framework, we will clarify the mathematical structure of the method and show how these expansions, together with the intermediate formulas, can be derived in a simple and natural~way.

\section{Preliminaries}
We can write Eq.~(1.5) in the form 
\begin{equation}
\frac{\partial}{\partial x}U(x,y;k)=  [2 i k J_3 - 2f(x) J_1] U(x,y;k)
\end{equation}
with
\begin{equation}
J_1 \equiv
-\frac{1}{2}
\begin{pmatrix}
0 & 1 \\
1 & 0 \\
\end{pmatrix},
\qquad
J_2 \equiv
\frac{1}{2}
\begin{pmatrix}
0 & -i \\
i & 0 \\
\end{pmatrix},
\qquad
J_3 \equiv 
\frac{1}{2}
\begin{pmatrix}
-1 & 0 \\
0 & 1 \\
\end{pmatrix}.
\end{equation}
The matrices $J_1, J_2, J_3$ satisfy the commutation relations
\begin{equation}
[J_1, J_2]=i J_3, \qquad
[J_2, J_3]=i J_1, \qquad
[J_3, J_1]=i J_2.
\end{equation}
We also define $J_+\equiv J_1 + i J_2$ and $J_- \equiv J_1 - i J_2$.
Then, $U(x,y;k)$ can be expressed as
\begin{equation}
U(x,y;k)=\exp[-R_r(x,y;k) J_+] \,
[\tau(x,y;k)]^{2 J_3}  \,
\exp[R_l(x,y;k) J_-],
\end{equation}
where $[\tau(x,y;k)]^{2 J_3} \equiv \exp\{[2 \log \tau(x,y;k)] J_3\}$.
Indeed, by using explicit forms (2.2), we can easily verify (2.4) by directly calculating the right-hand side as\cite{note2}
\begin{equation}
e^{-R_r J_+} \tau^{2 J_3} e^{R_l J_-} 
=
\begin{pmatrix}
1 & 0 \\
R_r & 1 \\
\end{pmatrix}
\begin{pmatrix}
\tau^{-1} & 0 \\
0 & \tau \\
\end{pmatrix}
\begin{pmatrix}
1 & - R_l \\
0 & 1 \\
\end{pmatrix}
=
\left(
\begin{array}{cc}
\alpha(k) & \beta(-k) \\
\beta(k) & \alpha(-k) \\
\end{array}
\right).
\end{equation}
Actually, we can prove (2.4) only by using commutation relations (2.3), without reference to explicit expressions (2.2) of $J_i$. 
This means that (2.4) is a representation-independent expression.\cite{algebraic}
Let $J_1,\,J_2,\,J_3$ be any set of operators satisfying commutation relations (2.3). Then, the solution of Eq.~(2.1) with the boundary condition $U(y,y)=1$ (identity operator) is given by (2.4).

Treating $f$ as a perturbation, we can derive from (1.5)--(1.7) the expressions of $\tau$, $R_l$, and $R_r$ as formal series in powers of $f$,
\begin{subequations}
\begin{align}
&\tau(x,y;k)=
e^{i k (x-y)}
- 
e^{i k (x-y)}
\int_y^x d z_1 \int_y^{z_1} d z_2 \,f(z_1) f(z_2) \,e^{2 i k(z_1-z_2)}
+ \cdots,
\\
&R_l(x,y;k)
=- \int_y^x d z_1 \,f(z_1)\,e^{2i k(z_1-y)}
\nonumber \\
& \qquad \qquad \qquad - \int_y^x d z_1 \int_y^{z_1} d z_2 \int_{z_2}^x d z_3\,
f(z_1) f(z_2) f(z_3) \,e^{2 i k(z_1 - z_2 + z_3 - y)} + \cdots,
\\
&R_r(x,y;k)
= \int_y^x d z_1\, f(z_1)\, e^{2 i k (x-z_1)}
\nonumber \\
& \qquad \qquad \qquad+\int_y^x d z_1 \int_{z_1}^x d z_2 \int_y^{z_2} d z_3\,
f(z_1)f(z_2)f(z_3)\, e^{2 i k (x - z_1 + z_2 - z_3)} + \cdots.
\end{align}
\end{subequations}
Equations (2.6) can be represented by diagrams as in Fig.~1(a).
These diagrams are to be interpreted according to the rules shown in Fig.~1(b).
%
%
%
\begin{figure}
\includegraphics[scale=0.62]{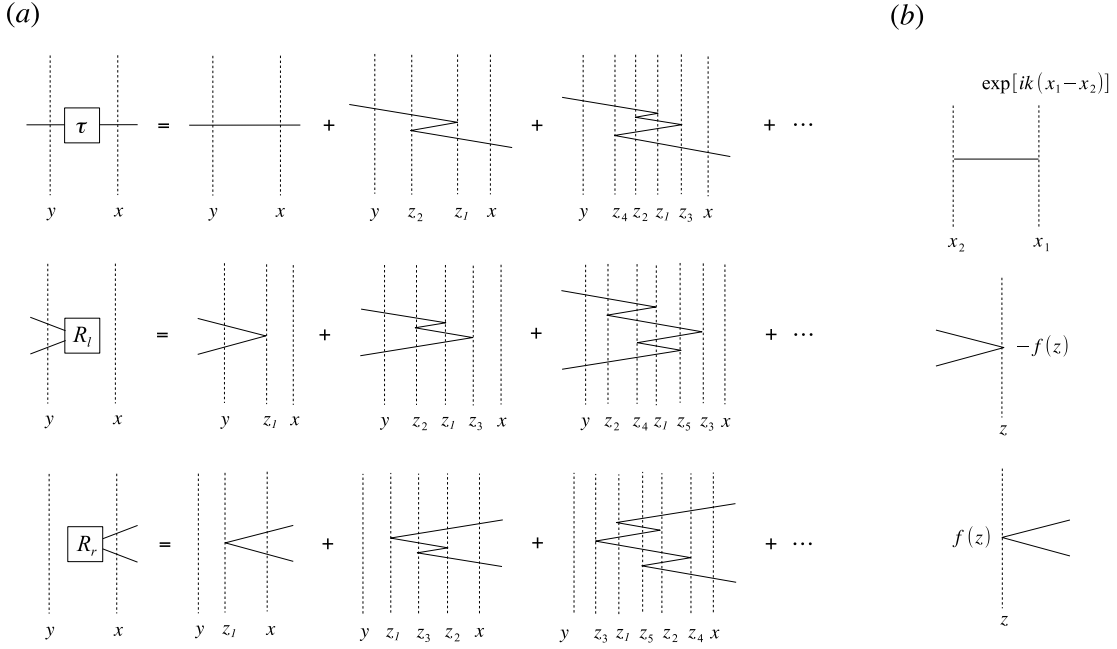}
\caption{
(a) Diagrammatic representation of Eqs.~(2.6).
(b) Rules for interpreting the diagrams.}
\end{figure}
Each line segment connecting the points $x_1$ and $x_2$ corresponds to $\exp[i k (x_2 - x_1)]$, and a factor $\pm f(z)$ is assigned to each turning point $z$. Integration over the positions of $z_i$ is implied in Fig.~1(a). (The vertical direction of the diagram does not have any meaning.)

Such diagrams are a useful tool for understanding the structure of the transmission and reflection coefficients in an intuitive way.
The series in (2.6) are formal, and we need not be concerned about their convergence.
Although we shall make use of diagrams in Sec.~III, the obtained results [such as (3.5) and (3.6)] are valid even if $f$ is not small and the series in (2.6) are not convergent.
These results can always be proved directly by using (1.7), without using series expressions (2.6).

The transmission and reflection coefficients become easier to handle by introducing an additional variable as follows.\cite{algebraic}
Suppose that $V(x)$ has a jump of magnitude $w$ at $x=x_0$, i.e., $\lim_{\epsilon \downarrow 0}\left[V(x_0 +\epsilon) - V(x_0 -\epsilon)\right]= w$. Then, $f(x)$ has a delta function singularity as
\begin{equation}
f(x)=-\frac{w}{2} \delta(x-x_0) + \cdots.
\end{equation}
From (1.5) and (2.7), we find that the matrix $U$ across the discontinuity takes the form
\begin{align}
\lim_{\epsilon \downarrow 0} U(x_0 + \epsilon, x_0-\epsilon;k)
&=\lim_{\epsilon \downarrow 0}\exp\left[
\int_{x_0-\epsilon}^{x_0 + \epsilon} d x
\begin{pmatrix}
-i k &  - \frac{w}{2}\delta(x-x_0) \\
- \frac{w}{2}\delta(x-x_0) & i k\\
\end{pmatrix}
\right]
\nonumber \\
&= \exp 
\begin{pmatrix}
0 & - w/2\\
- w/2 & 0\\
\end{pmatrix}
=
\begin{pmatrix}
\cosh \frac{w}{2}  &  -\sinh \frac{w}{2}\\
-\sinh \frac{w}{2} & \cosh \frac{w}{2}\\
\end{pmatrix}.
\end{align}
This matrix, which does not depend on $x_0$, describes the jump of the Fokker-Planck potential~$V$.
We multiply $U(x,y;k)$ [Eq.~(1.6)] from the left by this matrix and define
\begin{equation}
\left(
\begin{array}{cc}
\bar \alpha(x,y;w;k) & \bar \beta(x,y;w;-k) \\
\bar \beta(x,y;w;k) & \bar \alpha(x,y;w;-k) \\
\end{array}
\right)
\equiv 
\left(
\begin{array}{cc}
\cosh \frac{w}{2}  &  -\sinh \frac{w}{2}\\
-\sinh \frac{w}{2} & \cosh \frac{w}{2}\\
\end{array}
\right)
U(x,y;k).
\end{equation}
(The bar does not mean complex conjugation.) 
Using $\bar \alpha$ and $\bar \beta$ we define, as we did in (1.7), 
\begin{gather}
\bar \tau \equiv 
\frac{1}{\bar \alpha(x,y;w;k)},
\qquad
\bar R_l \equiv 
-\frac{\bar \beta(x,y;w;-k)}{\bar \alpha(x,y;w;k)},
\qquad
\bar R_r \equiv 
\frac{\bar \beta(x,y;w;k)}{\bar \alpha(x,y;w;k)}.
\end{gather}
We can interpret $\bar \tau$, $\bar R_l$, and $\bar R_r$ as the transmission and reflection coefficients that include the effect of a jump in $V(x)$ at the right endpoint of the interval. The additional variable $w$ corresponds to the magnitude of the jump.
It is convenient to introduce, instead of $w$, the new variable
\begin{equation}
\xi \equiv \tanh \frac{w}{2}
\end{equation}
and regard $\bar \tau$, $\bar R_r$, and $\bar R_l$ as functions of $\{x, y, \xi, k\}$. For simplicity, we shall often omit to write the argument $k$.
From (1.6), (1.7), (2.9), and (2.10) we have\cite{note2}
\begin{gather}
\bar \tau(x,y;\xi)
=\frac{\sqrt{1 - \xi^2}\, \tau(x,y)}{1-\xi R_r(x,y)},
\qquad
\bar R_l(x,y;\xi)
=R_l(x,y) + \frac{\xi \tau^2(x,y)}{1-\xi R_r(x,y)},
\nonumber \\
\bar R_r(x,y;\xi)=\frac{(1 - \xi^2) R_r(x,y)}{1-\xi R_r(x,y)} - \xi.
\end{gather}

\section{Structure of the generalized transmission and reflection coefficients}
Let us further generalize (2.12) and define
\begin{gather}
\hat T(x,y;\xi,\mu) \equiv \frac{\mu \tau(x,y)}{1 - \xi R_r(x,y)},
\qquad
\hat L(x,y;\xi) \equiv
R_l(x,y) + \frac{\xi \tau^2(x,y)}{1-\xi R_r(x,y)}, 
\nonumber \\
\hat R(x,y;\xi,\mu) \equiv
\frac{\mu^2 R_r(x,y)}{1 - \xi R_r(x,y)},
\end{gather}
where $\xi$ and $\mu$ are complex variables. 
Equations (2.12) can be written as
\begin{gather}
\bar \tau(x,y;\xi) = \hat T(x,y;\xi,\gamma),
\qquad
\bar R_l(x,y;\xi)= \hat L(x,y;\xi), 
\nonumber \\
\bar R_r(x,y;\xi)=\hat R(x,y;\xi,\gamma) - \xi,
\end{gather}
where
$\gamma \equiv \sqrt{1 - \xi^2}$ .
While $\xi$ and $\gamma$ are related by $\xi^2 + \gamma^2 =1$, the variables $\xi$ and $\mu$ in (3.1) are independent.
Moreover, we regard $\xi$ and $\mu$ as complex numbers.
The original $\tau$, $R_l$, and $R_r$ are recovered from $\hat T$, $\hat L$, and $\hat R$, respectively, by setting $\xi=0$ and $\mu=1$.

The reflection coefficient has the property that $\vert R_r \vert \leq 1$ for $\mathrm{Im}\,k \geq 0$. So, obviously $\hat T$, $\hat L$, and $\hat R$ are analytic functions of $\xi$ in the unit circle $\vert \xi \vert <1$.
The meaning of Eqs.~(3.1)  can be intuitively understood 
by expressing them as infinite series
\begin{align}
\hat T = \mu \tau  \sum_{n=0}^\infty (\xi R_r)^n,
\qquad
\hat L = R_l +\xi \tau^2 \sum_{n=0}^\infty (\xi R_r)^n, 
\qquad
\hat R =\mu^2 R_r \sum_{n=0}^\infty (\xi R_r)^n,
\end{align}
which are convergent for $\vert \xi \vert < 1$. 
These equations are illustrated in Fig.~2. 
%
%
%
\begin{figure}
\includegraphics[scale=0.57]{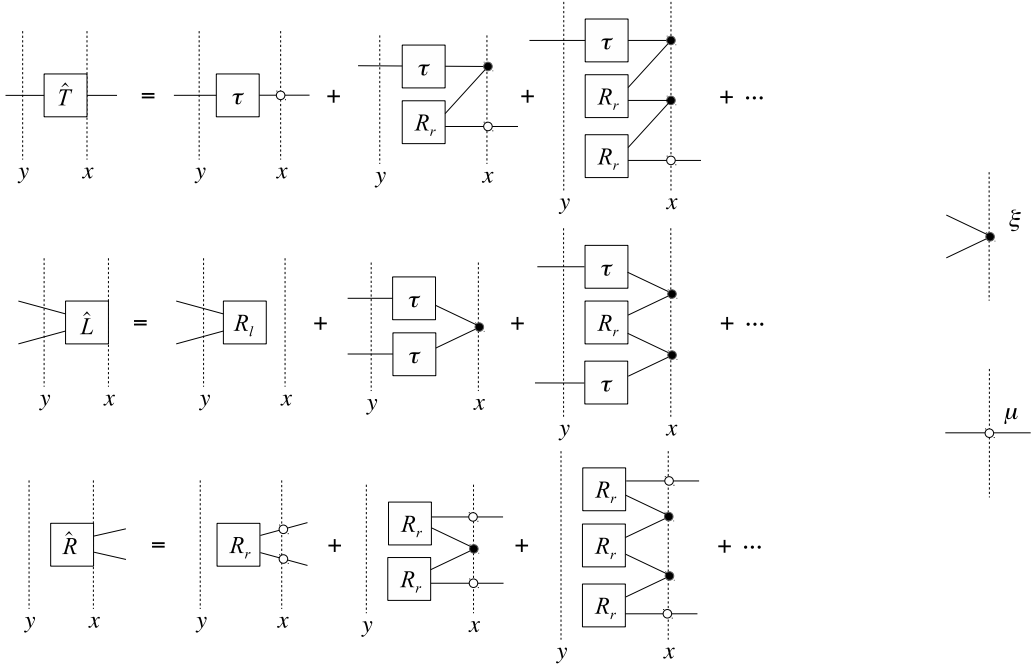}
\caption{
Schematic illustration of Eqs.~(3.3).
}
\end{figure}
It can be seen from the definition that $\tau(x,x)=1$, $R_r(x,x)=R_l(x,x)=0$. Hence,
\begin{equation}
\hat T(x,x;\xi,\mu)= \mu, \qquad
\hat L(x,x;\xi)=\xi, \qquad
\hat R(x,x;\xi,\mu)=0.
\end{equation}

Let $z \leq y \leq x$. In the diagrammatic representation, $\hat T(x,z)$, $\hat L(x,z)$, $\hat R(x,z)$ can be constructed from $\hat T(x,y)$, $\hat L(x,y)$, $\hat R(x,y)$ and $\tau(y,z)$, $R_l(y,z)$, $R_r(y,z)$ as shown in Fig.~3. 
%
%
%
\begin{figure}
\includegraphics[scale=0.67]{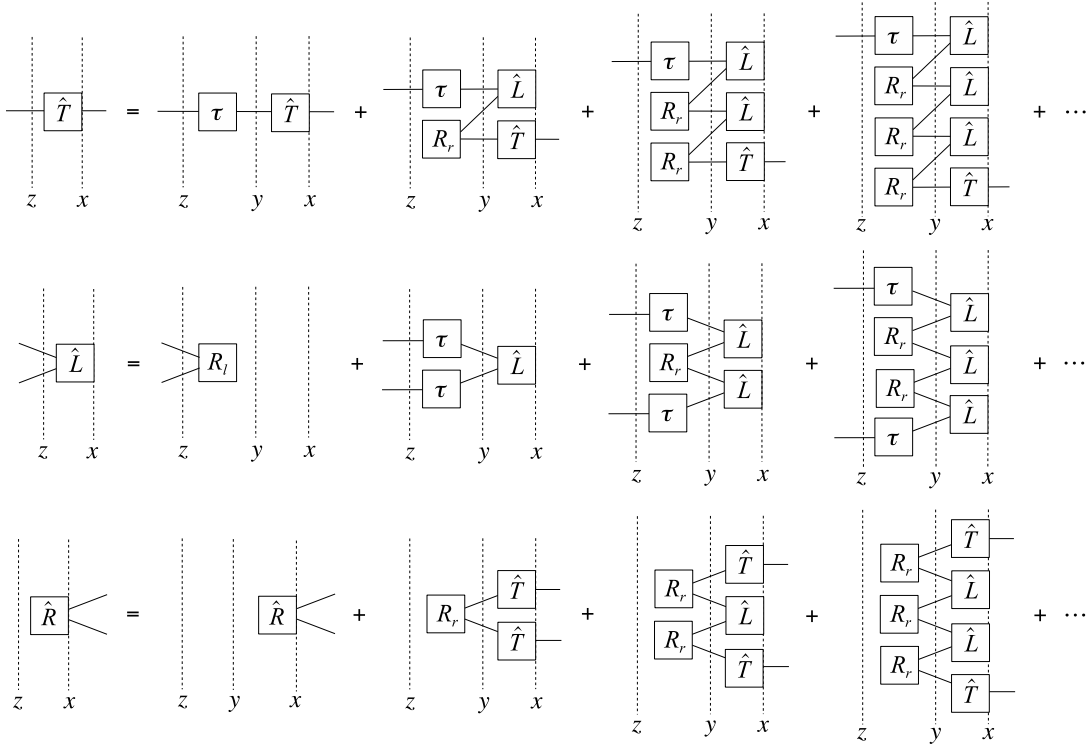}
\caption{
Construction of $\hat T(x,z)$, $\hat L(x,z)$, $\hat R(x,z)$ from $\hat T(x,y)$, $\hat L(x,y)$, $\hat R(x,y)$ and $\tau(y,z)$, $R_l(y,z)$, $R_r(y,z)$
[Eqs.~(3.5) and (3.6)].
}
\end{figure}
Comparing Fig.~3 with Fig.~2, we find that $\hat T(x,z)$ and $\hat L(x,z)$ are obtained from $\hat T(y,z;\xi,\mu)$ and $\hat L(y,z;\xi)$ by putting $\hat L(x,y)$ and $\hat T(x,y)$ in place of $\xi$ and $\mu$, respectively. That is,
\begin{equation}
\hat T(x,z;\xi,\mu)
= \hat T(y,z;\hat L(x,y;\xi),\hat T(x,y;\xi,\mu)), \qquad
\hat L(x,z;\xi)=\hat L(y,z;\hat L(x,y;\xi)).
\end{equation}
In the bottom row of Fig.~3, the first term on the right-hand side is exceptional and must be treated separately. So, the expression for $\hat R(x,z)$ includes an additional term as
\begin{equation}
\hat R(x,z;\xi,\mu)=\hat R(y,z;\hat L(x,y;\xi) ,\hat T(x,y;\xi,\mu)) + \hat R(x,y;\xi,\mu).
\end{equation}
Differentiating (3.5) and (3.6) with respect to $x$, and then setting $y = x$, we obtain
\begin{align}
&\frac{\partial}{\partial x}\hat T(x,z;\xi,\mu)
=
\frac{\partial}{\partial \xi} \hat T(x,z;\xi,\mu)
\left.
\frac{\partial}{\partial x} \hat L(x,y;\xi)
\right\vert_{y=x}
 + 
\frac{\partial}{\partial \mu} \hat T(x,z;\xi,\mu)
\left.
\frac{\partial}{\partial x} \hat T(x,y;\xi,\mu)
\right\vert_{y=x} ,
\nonumber \\
&\frac{\partial}{\partial x}\hat L(x,z;\xi)
=
\frac{\partial}{\partial \xi} \hat L(x,z;\xi)
\left.
\frac{\partial}{\partial x} \hat L(x,y;\xi)
\right\vert_{y=x},
\nonumber \\
&\frac{\partial}{\partial x}\hat R(x,z;\xi,\mu)
=
\frac{\partial}{\partial \xi} \hat R(x,z;\xi,\mu)
\left.
\frac{\partial}{\partial x} \hat L(x,y;\xi)
\right\vert_{y=x} 
+ 
\frac{\partial}{\partial \mu} \hat R(x,z;\xi,\mu)
\left.
\frac{\partial}{\partial x} \hat T(x,y;\xi,\mu)
\right\vert_{y=x} 
\nonumber \\*
& \qquad \qquad \qquad\qquad
+\left.
\frac{\partial}{\partial x} \hat R(x,y;\xi,\mu)
\right\vert_{y=x},
\end{align}
where we have used (3.4).
From Fig.~4, we find that (see Appendix~A)
%
%
%
\begin{figure}
\hspace{-1cm}
\includegraphics[scale=0.55]{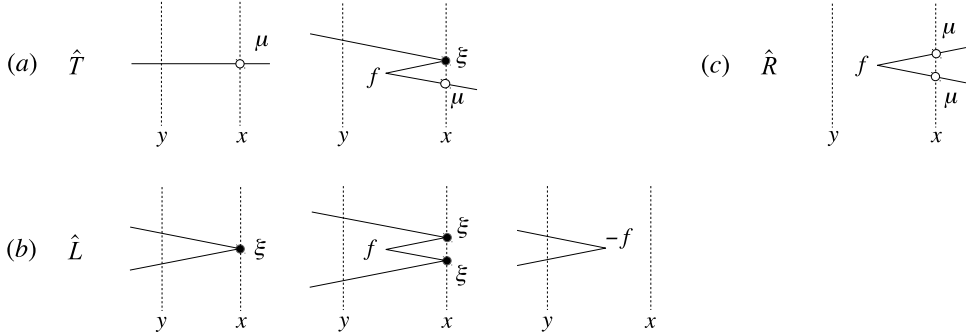}
\caption{
Explanation of Eqs.~(3.8) using diagrams.
}
\end{figure}
\begin{align}
&\left.
\frac{\partial}{\partial x} \hat T(x,y;\xi,\mu) 
\right\vert_{y=x} 
= i k \mu + \mu \xi f(x),
\nonumber \\
&\left.
\frac{\partial}{\partial x} \hat L(x,y;\xi)
\right\vert_{y=x} 
=2 i k \xi + (\xi^2 -1) f(x),
\qquad
\left.
\frac{\partial}{\partial x} \hat R(x,y;\xi,\mu) 
\right\vert_{y=x} 
= \mu^2 f(x).
\end{align}
Equations (3.7) can be simply expressed by introducing the differential operators
\begin{equation}
\mathcal{J}_1 \equiv \frac{1}{2} (1-\xi^2) \frac{\partial}{\partial \xi}
-\frac{1}{2} \xi \mu \frac{\partial}{\partial \mu},
\qquad
\mathcal{J}_2 \equiv \frac{i}{2} (1+\xi^2) \frac{\partial}{\partial \xi}
+\frac{i}{2} \xi \mu \frac{\partial}{\partial \mu},
\qquad
\mathcal{J}_3 \equiv \xi \frac{\partial}{\partial \xi} + \frac{1}{2}\mu \frac{\partial}{\partial \mu}.
\end{equation}
These operators satisfy the same commutation relations as (2.3), i.e., $[\mathcal{J}_1, \mathcal{J}_2]= i \mathcal{J}_3$, $[\mathcal{J}_2, \mathcal{J}_3]= i \mathcal{J}_1$, $[\mathcal{J}_3$, $\mathcal{J}_1]= i \mathcal{J}_2$.
Using (3.8) and (3.9), we can write (3.7) as
\begin{subequations}
\begin{align}
&\frac{\partial}{\partial x} \hat T(x,y;\xi,\mu)
=[2 i k \mathcal{J}_3 - 2f(x) \mathcal{J}_1 ] \hat T(x,y;\xi,\mu), 
\\
&\frac{\partial}{\partial x} \hat L(x,y;\xi)
=[2 i k \mathcal{J}_3 - 2f(x) \mathcal{J}_1 ] \hat L(x,y;\xi) ,
\\
&\frac{\partial}{\partial x} \hat R(x,y;\xi,\mu)
=[2 i k \mathcal{J}_3 - 2f(x) \mathcal{J}_1 ] \hat R(x,y;\xi,\mu) 
+ \mu^2 f(x).
\end{align}
\end{subequations}
(We have replaced $z$ by $y$.) 
Note that (3.10a) and (3.10b) have the same form as (2.1).

In the same way as (2.1), we define the evolution operator $\mathcal{U}(x,y)$, which is an operator acting on functions of $\xi$ and $\mu$, as the solution of the differential equation
\begin{equation}
\frac{\partial}{\partial x} \mathcal{U}(x,y)
=[2 i k \mathcal{J}_3 - 2 f(x) \mathcal{J}_1] \, \mathcal{U}(x,y)
\end{equation}
with the boundary condition $\mathcal{U}(y,y)=1$ (identity operator).
According to general formula (2.4) (which is representation independent), this $\mathcal{U}$ can be expressed as
\begin{equation}
\mathcal{U}(x,y)
=\exp[-R_r(x,y) \mathcal{J}_+][\tau(x,y)]^{2 \mathcal{J}_3} \exp[R_l(x,y) \mathcal{J}_-],
\end{equation}
where
\begin{equation}
\mathcal{J}_+
\equiv \mathcal{J}_1 + i \mathcal{J}_2 
=-\xi^2 \frac{\partial}{\partial \xi} - \xi \mu \frac{\partial}{\partial \mu}, 
\qquad
\mathcal{J}_- 
\equiv \mathcal{J}_1 - i \mathcal{J}_2 
=\frac{\partial}{\partial \xi}.
\end{equation}
When acting on $\mu$-independent functions of $\xi$, the operators $\mathcal{J}_\pm$ and $\mathcal{J}_3$ reduce to
\begin{equation}
\mathcal{J}_+^{(0)}= -\xi^2 \frac{\partial}{\partial \xi}, 
\qquad
\mathcal{J}_-^{(0)} = \frac{\partial}{\partial \xi}, 
\qquad
\mathcal{J}_3^{(0)} = \xi \frac{\partial}{\partial \xi}.
\end{equation}
It is well known that $\mathcal{J}_3^{(0)}$, $\mathcal{J}_-^{(0)}$, and $\mathcal{J}_+^{(0)}$ are, respectively, the generators of dilatation, translation, and special conformal transformation,\cite{cardy} which means that
\begin{align}
&\exp(c \mathcal{J}_3^{(0)}) g(\xi)=g(e^c \xi),
\qquad
\exp(c \mathcal{J}_-^{(0)})g(\xi)=g(\xi + c),
\nonumber \\
&\exp(c \mathcal{J}_+^{(0)})g(\xi)=g\Bigl(\frac{\xi}{1 + c \xi}\Bigr)
\end{align}
for an arbitrary constant $c$.
These relations can be generalized to the $\mu$-dependent case as
\begin{align}
&\exp(c \mathcal{J}_3) g(\xi,\mu)=g(e^c \xi, e^{c/2}\mu),
\qquad
\exp(c \mathcal{J}_-)g(\xi,\mu)=g(\xi + c, \mu),
\nonumber \\
&\exp(c \mathcal{J}_+)g(\xi,\mu)=g\Bigl(\frac{\xi}{1 + c \xi}, \frac{\mu}{1 + c \xi}\Bigr).
\end{align}
The first equation of (3.16) easily follows from $\exp(c \mathcal{J}_3)= \exp(\frac{c}{2}\mu \frac{\partial}{\partial \mu} )\exp(c \xi \frac{\partial}{\partial \xi})$. The second equation of (3.16) is a trivial extension of (3.15). Defining $\zeta \equiv \xi$ and $\alpha \equiv \mu/\xi$, we have
\begin{equation}
\exp(c \mathcal{J}_+)g(\xi,\mu)
=\exp\left(-c \,\zeta^2\frac{\partial}{\partial \zeta}\right) g(\zeta, \alpha \zeta).
\end{equation}
The third equation of (3.16) follows from (3.17) and the third equation of (3.15).

Using (3.12) and (3.16), we can see that $\mathcal{U}$ acts on a function $g(\xi,\mu)$ as
\begin{align}
\mathcal{U}(x,y)g(\xi,\mu)
&=\exp(-R_r \mathcal{J}_+)\tau^{2 \mathcal{J}_3} \exp(R_l\mathcal{J}_-)g(\xi,\mu)
=\exp(-R_r \mathcal{J}_+)\tau^{2 \mathcal{J}_3}  g(\xi + R_l, \mu)
\nonumber \\*
&=\exp(-R_r \mathcal{J}_+)g(\tau^2 \xi + R_l, \tau \mu)
=g\Bigl(\frac{\tau^2 \xi}{1-R_r \xi} + R_l, \frac{\tau \mu}{1-R_r \xi}\Bigr).
\end{align}
In view of definition (3.1), this expression reads
\begin{equation}
\mathcal{U}(x,y)g(\xi,\mu) = g(\hat L(x,y;\xi), \hat T(x,y;\xi, \mu)).
\end{equation}
Thus, applying the operator $\mathcal{U}(x,y)$ to a function $g(\xi,\mu)$ amounts to replacing $\xi$ and $\mu$ by $\hat L(x,y;\xi)$ and $\hat T(x,y;\xi,\mu)$, respectively.
In particular,
\begin{equation}
\mathcal{U}(x,y) \mu= \hat T(x,y;\xi,\mu), \qquad
\mathcal{U}(x,y) \xi= \hat L(x,y;\xi).
\end{equation}
When the function $g$ in (3.19) has one or more spatial variables in addition to $\xi$ and $\mu$, these variables remain unchanged on the right-hand side of (3.19). For example,
\begin{equation}
\mathcal{U}(x,y)g(y,\xi,\mu) = g(y,\hat L(x,y;\xi), \hat T(x,y;\xi, \mu)).
\end{equation}
As can be seen from the definition, the evolution operator has the property
\begin{equation}
\mathcal{U}(x,y)\,\mathcal{U}(y,z)=\mathcal{U}(x,z).
\end{equation}
From (3.22), it follows that $\mathcal{U}(y,x)=[\mathcal{U}(x,y)]^{-1}$. From (3.22) and (3.20), we have
\begin{equation}
\mathcal{U}(x,y) \hat T(y,z;\xi,\mu)= \hat T(x,z;\xi,\mu),
\qquad
\mathcal{U}(x,y) \hat L(y,z;\xi)= \hat L(x,z;\xi).
\end{equation}
This is none other than Eqs.~(3.5) rewritten using $\mathcal{U}$.
Similarly, (3.6) can be rewritten as
\begin{equation}
\mathcal{U}(x,y) \hat R(y,z;\xi,\mu)=\hat R(x,z;\xi,\mu) - \hat R(x,y;\xi,\mu).
\end{equation}

\section{Solution of the inhomogeneous equation}
Unlike (3.10a) and (3.10b), Eq.~(3.10c) is inhomogeneous. We define
\begin{equation}
\mathcal{A} \equiv \frac{\partial}{\partial x} + 2 f(x) \mathcal{J}_1,
\qquad
\mathcal{B}\equiv \mathcal{J}_3,
\end{equation}
and write (3.10c) as
\begin{equation}
(\mathcal{A} - 2 i k \mathcal{B}) \hat R(x,y;\xi,\mu)=\mu^2 f(x).
\end{equation}
More generally than (4.2), let us consider the inhomogeneous differential equation
\begin{equation}
(\mathcal{A} - 2 i k \mathcal{B}) h(x,\xi,\mu)= g(x,\xi,\mu),
\end{equation}
where $h$ is an unknown function and $g$ is a given function of $x$, $\xi$, and $\mu$. 

We can rewrite (3.22) as $\mathcal{U}(z,x)\mathcal{U}(x,y)=\mathcal{U}(z,y).$ Differentiating both sides of this equation with respect to $x$, and then setting $y=x$ and using $\mathcal{U}(x,x)=1$, we have 
\begin{equation}
\frac{\partial}{\partial x} \mathcal{U}(z,x)=
\mathcal{U}(z,x)[- 2 i k\mathcal{J}_3 + 2f(x) \mathcal{J}_1].
\end{equation}
Therefore, for arbitrary $z$,
\begin{equation}
\frac{\partial}{\partial x}[\mathcal{U}(z,x) h(x,\xi,\mu)]
=
\mathcal{U}(z,x)
\left[\frac{\partial}{\partial x}- 2 i k\mathcal{J}_3 + 2f(x) \mathcal{J}_1\right]
h(x,\xi,\mu).
\end{equation}
Substituting (4.3), and interchanging $x$ and $z$, we rewrite (4.5) as
\begin{equation}
\frac{\partial}{\partial z}[\,\mathcal{U}(x,z) h(z,\xi,\mu)]
=\mathcal{U}(x,z) g(z,\xi,\mu).
\end{equation}

To solve (4.2) or (4.3), we need a boundary condition. 
When $y$ is finite, the boundary condition for $\hat R(x,y;\xi,\mu)$ is $\hat R(y,y;\xi,\mu)=0$ for any $\xi$ and $\mu$ [see Eq.~(3.4)].
Suppose that the solution $h$ of (4.3) satisfies the boundary condition $h(y,\xi,\mu)=0$, with a fixed number~$y$, for any $\xi$ and $\mu$. 
Integrating (4.6) from $z=y$ to $z=x$ gives
\begin{equation}
h(x, \xi, \mu)=\int_y^x d z\, \mathcal{U}(x,z) g(z,\xi,\mu),
\end{equation}
where we have used $\mathcal{U}(x,y) h(y,\xi,\mu)=h(y,\hat L(x,y;\xi), \hat T(x,y;\xi,\mu))=0$.
Hence,
\begin{equation}
\hat R(x,y;\xi,\mu)=\int_y^x d z\,\mathcal{U}(x,z) \mu^2 f(z)
= \int_y^x d z\,[\hat T(x,z;\xi,\mu)]^2 f(z).
\end{equation}

In this paper, we are interested in reflection coefficients for semi-infinite intervals.
Corresponding to (1.8), the definition of $\hat R$ for a semi-infinite interval is
\begin{equation}
\hat R(x,-\infty;\xi,\mu) = \lim_{y \to -\infty} \hat R(x,y;\xi,\mu).
\end{equation}
Provided that the integral in (4.8) is convergent in the limit $y \to -\infty$, we have
\begin{equation}
\hat R(x,-\infty;\xi,\mu)
= \int_{-\infty}^x d z\,[\hat T(x,z;\xi,\mu)]^2 f(z).
\end{equation}
It is also possible to derive (4.10) directly by considering Eq.~(4.2) with $y = -\infty$. When dealing with $\hat R(x,-\infty)$, we need to be careful about the boundary condition. 
The boundary condition for $\hat R(x,-\infty)$ is not $\lim_{x\to -\infty} \hat R(x,-\infty)=0$. (This does not necessarily hold.)
Setting $z=-\infty$ in (3.24) and using (4.9), we find that
\begin{equation}
\lim_{y \to -\infty}\mathcal{U}(x,y) \hat R(y,-\infty;\xi,\mu)=0,
\end{equation}
where $x$ is arbitrary. This is the correct boundary condition for $\hat R(x,-\infty)$.
Let us assume that the solution $h$ of equation~(4.3) satisfies the same boundary condition as (4.11),
\begin{equation}
\lim_{z \to -\infty} \mathcal{U}(x, z)h(z,\xi,\mu)=0,
\end{equation}
for any $\xi$ and $\mu$. Integrating (4.6) from $z=-\infty$ to $z=x$, and using (4.12), we obtain
\begin{equation}
h(x,\xi,\mu)
=\int_{-\infty}^x d z\,  \mathcal{U}(x,z) g(z,\xi,\mu)
=\int_{-\infty}^x d z\, g(z,\hat L(x,z;\xi),\hat T(x,z;\xi,\mu)).
\end{equation}

In the following analysis of $\hat R(x,-\infty)$, we will mainly deal with functions that satisfy boundary condition (4.12). 
So, let us restrict the domain of the operator $(\mathcal{A} - 2 i k \mathcal{B})$ to functions satisfying (4.12). Then, as shown above, the inverse of $(\mathcal{A} - 2 i k \mathcal{B})$ is given by
\begin{equation}
(\mathcal{A} - 2 i k \mathcal{B})^{-1} g(x,\xi,\mu)
=\int_{-\infty}^x d z\, g(z,\hat L(x,z;\xi),\hat T(x,z;\xi,\mu)).
\end{equation}
Using this inverse, the solution of (4.3) is obtained as $h=(\mathcal{A} - 2 i k \mathcal{B})^{-1} g$.
In other words, the solution of (4.3) is given by (4.14) if we assume that $h$ satisfies (4.12). Since $\hat R(x,-\infty)$ satisfies equation (4.2) and boundary condition (4.11), we can express it as
\begin{equation}
\hat R(x, -\infty;\xi, \mu)=(\mathcal{A} - 2 i k \mathcal{B})^{-1} \mu^2 f(x).
\end{equation}

\section{Transformation of the variables}
The operator $\mathcal{J}_3$ is diagonalized in the basis of its eigenfunctions $\mu^n \xi^m$.
When dealing with the low-energy expansion, it is convenient (as we shall see in Sec.~VII) to use a basis in which $\mathcal{J}_1$, rather than $\mathcal{J}_3$, is diagonalized. 
To find such a basis, we note that $\mathcal{J}_1$ and $\mathcal{J}_3$ are related by a rotation of angle $\pi/2$ around the $2$-axis, which is represented by the operator $\exp(-\frac{1}{2} i \pi \mathcal{J}_2)$. It follows from commutation relations (2.3) that
\begin{equation}
\exp \Bigl(\frac{i \pi}{2}\mathcal{J}_2 \Bigr)
\mathcal{J}_1
\exp \Bigl(-\frac{i \pi}{2}\mathcal{J}_2 \Bigr)
=\mathcal{J}_3,
\qquad
\exp \Bigl(\frac{i \pi}{2}\mathcal{J}_2 \Bigr)
\mathcal{J}_3
\exp \Bigl(-\frac{i \pi}{2}\mathcal{J}_2 \Bigr)
=-\mathcal{J}_1.
\end{equation}
Applying this rotation to $\xi$ and $\mu$ gives (see Appendix~B)
\begin{equation}
\tilde \xi \equiv \exp \Bigl(-\frac{i \pi}{2}\mathcal{J}_2 \Bigr)\xi =\frac{1+\xi}{1-\xi},
\qquad
\tilde \mu \equiv \exp\Bigl(-\frac{i \pi}{2}\mathcal{J}_2 \Bigr)\mu= \frac{\sqrt{2} \mu}{1-\xi}.
\end{equation}
Let $\tilde{\mathcal{J}}_1,\,\tilde{\mathcal{J}}_2,\,\tilde{\mathcal{J}}_3$ be defined by (3.9) with $\xi$ and $\mu$ replaced by $\tilde \xi$ and $\tilde \mu$,
\begin{equation}
\tilde{\mathcal{J}}_1 \equiv
\frac{1}{2} (1-\tilde \xi^2) \frac{\partial}{\partial \tilde \xi}
-\frac{1}{2} \tilde \xi \tilde \mu \frac{\partial}{\partial \tilde \mu},
\qquad
\tilde{\mathcal{J}}_2 \equiv
\frac{i}{2} (1+\tilde \xi^2) \frac{\partial}{\partial \tilde \xi}
+\frac{i}{2} \tilde \xi \tilde \mu \frac{\partial}{\partial\tilde  \mu},
\qquad
\tilde{\mathcal{J}}_3 \equiv \tilde \xi \frac{\partial}{\partial \tilde \xi} + \frac{1}{2}\tilde \mu \frac{\partial}{\partial \tilde \mu}.
\end{equation}
If we define the variables $w$ and $a$ by $\tilde \xi \equiv e^w$ and $\tilde \mu \equiv \sqrt{2} a e^{w/2}$, expressions (5.3) become
\begin{gather}
\tilde{\mathcal{J}}_1= -(\sinh w) \frac{\partial}{\partial w} - \frac{1}{2} (\cosh w) a \frac{\partial}{\partial a},
\qquad
\tilde{\mathcal{J}}_2=  i (\cosh w) \frac{\partial}{\partial w} + \frac{i}{2} (\sinh w) a \frac{\partial}{\partial a},
\nonumber \\
\tilde{\mathcal{J}}_3= \frac{\partial}{\partial w}.
\end{gather}
Thus, $\tilde{\mathcal{J}}_3$ takes a simple form.
The original variables $\xi$ and $\mu$ are related to $w$ and $a$ as
\begin{equation}
\xi = \tanh \frac{w}{2}, \qquad \mu=\frac{a}{\cosh \frac{w}{2}}
\qquad \mbox{or} \qquad
w=\log \frac{1+ \xi}{1-\xi}, \qquad a=\frac{\mu}{\sqrt{1-\xi^2}}.
\end{equation}
The first equation of (5.5) has the same form as (2.11). The variable $w$ has the same meaning as in Sec.~II when restricted to real numbers. (Now $\xi$ and $w$ are complex numbers.)

From (5.1), it follows that $\mathcal{J}_1 = \tilde{\mathcal{J}}_3$ and $\mathcal{J}_3 = - \tilde{\mathcal{J}}_1$. 
Using (5.4) in (4.1) gives
\begin{equation}
\mathcal{A}=\frac{\partial}{\partial x} + 2 f(x) \frac{\partial}{\partial w},
\qquad
\mathcal{B}=(\sinh w) \frac{\partial}{\partial w} + \frac{1}{2} (\cosh w) a \frac{\partial}{\partial a}.
\end{equation}
To simplify the expression for $\mathcal{A}$, let us define
\begin{equation}
W\equiv w + V(x).
\end{equation}
For any function $g(x,w)=g(x,W-V(x))$, we have the relations
\begin{equation}
\left(\frac{\partial g}{\partial x}\right)_W
=\left(\frac{\partial g}{\partial x}\right)_w + 2 f(x)\left(\frac{\partial g}{\partial w}\right)_x,
\qquad
\left(\frac{\partial g}{\partial W}\right)_x = \left(\frac{\partial g}{\partial w}\right)_x,
\end{equation}
where $(\partial g/\partial x)_W$ and $(\partial g/\partial x)_w$ denote the partial derivatives with fixed $W$ and fixed $w$, respectively. 
Let us now change the set of independent variables from $\{x,w,a\}$ to $\{x,W,a\}$. 
We assume that the operators $\mathcal{A}$ and $\mathcal{B}$ act on functions of $x$, $W$, and $a$. 
From (5.8), we see that the expressions for $\mathcal{A}$ and $\mathcal{B}$ become
\begin{equation}
\mathcal{A}= \frac{\partial}{\partial x},
\qquad
\mathcal{B}= \sinh [W-V(x)] \frac{\partial}{\partial W} 
+ \frac{1}{2} \cosh [W-V(x)]\, a \frac{\partial}{\partial a}.
\end{equation}
Equations (3.21) and (4.14) can be written in the variables $\{x,W,a\}$ as
\begin{equation}
\mathcal{U}(x,z) g(z,W,a) = g(z,\bar \omega(x,z;W), \bar A(x,z;W,a)),
\end{equation}
\begin{equation}
(\mathcal{A}- 2 i k \mathcal{B})^{-1} g(x,W,a)
= \int_{-\infty}^x d z\,g(z,\bar \omega(x,z;W), \bar A(x,z;W,a)),
\end{equation}
where 
\begin{equation}
\bar \omega(x,z;W)  \equiv \log \frac{1+ \hat L(x,z;W)}{1- \hat L(x,z;W)} + V(z), 
\qquad \!\!\!
\bar A(x,z;W,a) \equiv \frac{\hat T(x,z;W,a)}{\sqrt{ 1 - \hat L^2(x,z;W) }}.
\end{equation}
Expressions (5.12) come from the last two equations of (5.5) and $W=w+V(z)$.
[Here, $W$ is not $w+V(x)$ but $w+V(z)$, since this $W$ appears in $g(z,W,a)$.]

In (5.12), the functions $\hat T$ and $\hat L$ are written with the variables $W$ and $a$. For functions which have two spatial variables such as $\hat T$ or $\hat L$, the variable $W$ is associated with the first spatial variable, which corresponds to the right endpoint of the interval. For example, in $\hat T(x_1,x_2;W,a)$, the definition of $W$ is $W=w+V(x_1)$. 
From (3.1), (5.5), and (5.7), we have
\begin{gather}
\hat T(x,y;W,a)=\frac{a\, \gamma(x,W)\, \tau(x,y)}
{1 - \xi(x,W)\, R_r(x,y)},
\qquad
\hat L(x,y;W) = R_l(x,y) + 
\frac{\xi(x,W)\, \tau^2(x,y)}
{1 - \xi(x,W)\,R_r(x,y)},
\nonumber \\
\hat R(x,y;W,a)=\frac{a^2\, \gamma^2(x,W)\,R_r(x,y)}
{1 - \xi(x,W)\, R_r(x,y)},
\end{gather}
where
\begin{equation}
\xi(x,W) = \tanh\frac{W-V(x)}{2},
\qquad
\gamma(x,W) = \sqrt{1 - \xi^2} = \mathrm{sech} \frac{W-V(x)}{2}.
\end{equation}
Similarly, we can rewrite Eqs.~(2.12) using the variables $\{x,y,W\}$ as 
\begin{gather}
\bar \tau(x,y;W)
=\frac{\gamma(x,W)\, \tau(x,y)}{1-\xi(x,W)R_r(x,y)},
\qquad
\bar R_l(x,y;W)
=R_l(x,y) + \frac{\xi(x,W) \tau^2(x,y)}{1-\xi(x,W)R_r(x,y)},
\nonumber \\
\bar R_r(x,y;W)=\frac{\gamma^2(x,W) R_r(x,y)}{1-\xi(x,W)R_r(x,y)} - \xi(x,W),
\end{gather}
Comparing (5.13) with (5.15), we find
\begin{align}
&\hat T(x,y;W,a)=a \bar \tau(x,y;W),
\qquad
\hat L(x,y;W) = \bar R_l(x,y;W),
\nonumber \\
&\hat R(x,y;W,a) = a^2 \left[\bar R_r(x,y;W) + \xi(x,W) \right].
\end{align}
The variable $W$ is a complex number. As noted before, $\hat T$, $\hat L$, and $\hat R$ are analytic functions of $\xi$ in the unit circle $\vert \xi \vert <1$. 
Since $V(x)$ is a real number, $\vert \xi(x,W)\vert <1$ corresponds to $\vert \mathrm{Im}\,W \vert < \pi/2$ irrespective of $x$. 
Therefore, $\hat T$, $\hat L$, and $\hat R$ are analytic functions of $W$ in the strip $\vert \mathrm{Im}\,W \vert < \pi/2$.

\section{Expressions in eigenspaces of $\boldsymbol{\mu \partial/\partial \mu}$}

The operators $\tilde{\mathcal{J}}_1, \, \tilde{\mathcal{J}}_2, \, \tilde{\mathcal{J}}_3$,  
as well as $\mathcal{J}_1, \, \mathcal{J}_2, \, \mathcal{J}_3$, satisfy the same commutation relations as (2.3). So, they constitute a representation of the Lie algebra $sl(2,{\bf C})$,
 where the representation space consists of analytic 
functions of $w$ and $a$ (or $\xi$ and $\mu$).  This representation is reducible.
All these operators commute with the operator defined by
\begin{equation}
\mathcal{N} \equiv \mu \frac{\partial}{\partial \mu} = a \frac{\partial}{\partial a} .
\end{equation}
When $w$ and $a$ are taken as independent variables, the eigenspace of $\mathcal{N}$ with eigenvalue $\nu$ consists of functions of the form $a^\nu g(w)$. In this eigenspace, $\tilde{\mathcal{J}}_1$ and $\tilde{\mathcal{J}}_2$ [Eqs.~(5.4)] reduce to
\begin{equation}
\tilde{\mathcal{J}}_1^{(\nu)} \equiv - \sinh w \frac{\partial}{\partial w} - \frac{\nu}{2} \cosh w,
\qquad 
\tilde{\mathcal{J}}_2^{(\nu)} \equiv  i \cosh w \frac{\partial}{\partial w} + i \frac{\nu}{2} \sinh w,
\end{equation}
which act on functions of $w$ alone.
That is to say,
\begin{equation}
\tilde{\mathcal{J}}_i \,a^\nu g(w)=a^\nu \tilde{\mathcal{J}}_i^{(\nu)} g(w)
\qquad
(i=1,2,3),
\end{equation}
where we have also defined $\tilde{\mathcal{J}}_3^{(\nu)} \equiv \partial/\partial w$.
The operators $\tilde{\mathcal{J}}_1^{(\nu)}$, $\tilde{\mathcal{J}}_2^{(\nu)}$, and $\tilde{\mathcal{J}}_3^{(\nu)}$ satisfy the same commutation relations as (2.3). 
Thus, we have a different representation of $sl(2,{\bf C})$ for each $\nu$.
[In this representation, the Casimir operator is $-\frac{\nu}{2} (1 - \frac{\nu}{2} )$ times the unit operator.] 

When the spatial variable $x$ is included and the independent variables are changed from $\{x, w, a\}$ to $\{x, W, a\}$, the eigenspace of $\mathcal{N}$ with eigenvalue $\nu$ consists of functions of the form $a^\nu g(x,W)$. 
In each eigenspace, the operators $\mathcal{A}$ and $\mathcal{B}$ [Eqs.~(5.9)] act as
\begin{equation}
\mathcal{A}\, a^\nu g(x,W) = a^\nu \mathcal{A}^{(\nu)} g(x,W), 
\qquad
\mathcal{B}\, a^\nu g(x,W) = a^\nu \mathcal{B}^{(\nu)} g(x,W),
\end{equation}
\begin{equation}
\mathcal{A}^{(\nu)} \equiv \frac{\partial}{\partial x}, \qquad
\mathcal{B}^{(\nu)} \equiv \sinh [W-V(x)] \frac{\partial}{\partial W} + \frac{\nu}{2} \cosh [W-V(x)].
\end{equation}
The inverses of $(\mathcal{A} - 2 i k \mathcal{B})$ and $(\mathcal{A}^{(\nu)} - 2 i k \mathcal{B}^{(\nu)})$ satisfy the corresponding relation
\begin{align}
(\mathcal{A} - 2 i k \mathcal{B})^{-1} a^\nu g(x,W) = a^\nu (\mathcal{A}^{(\nu)} - 2 i k \mathcal{B}^{(\nu)})^{-1}g(x,W).
\end{align}
Using (5.11) for the left-hand side of (6.6), and inserting (5.16), we find
\begin{equation}
(\mathcal{A}^{(\nu)} - 2 i k \mathcal{B}^{(\nu)})^{-1}g(x,W)
= \int_{-\infty}^x d z\,
\left[
\frac{\bar \tau^2(x,z;W)}{1 - \bar R_l^2(x,z;W)}
\right]^{\nu/2} g(z,\bar \omega(x,z;W)).
\end{equation}

If $\{x,\xi,\mu\}$ are taken as the independent variables instead of $\{x, W, a\}$, the eigenspace of $\mathcal{N}$ consists of functions of the form $\mu^\nu g(x,\xi)$. 
From (4.1) and (3.9), we have
\begin{equation}
\mathcal{A}\, \mu^\nu g(x,\xi) = \mu^\nu \mathcal{A}_{(\nu)} g(x,\xi), 
\qquad
\mathcal{B}\, \mu^\nu g(x,\xi) = \mu^\nu \mathcal{B}_{(\nu)} g(x,\xi),
\end{equation}
\begin{align}
\mathcal{A}_{(\nu)} \equiv
\frac{\partial}{\partial x} + f(x) \left[(1 - \xi^2) \frac{\partial}{\partial \xi} - \nu \xi \right],
\qquad
\mathcal{B}_{(\nu)} \equiv \xi \frac{\partial}{\partial \xi} + \frac{\nu}{2}. 
\end{align}
As shown in (5.5), 
$\mu = \gamma a$ with $\gamma \equiv \sqrt{1-\xi^2}$.
Comparing (6.4) with (6.8), we can see that
\begin{equation}
\mathcal{A}_{(\nu)} = \gamma^{-\nu} \mathcal{A}^{(\nu)}\, \gamma^\nu, 
\qquad
\mathcal{B}_{(\nu)} = \gamma^{-\nu} \, \mathcal{B}^{(\nu)} \gamma^\nu.
\end{equation}

The number $\nu$ corresponds to the number of white circles in each diagram in Fig.~2.
The functions $\hat T$, $\hat L$, and $\hat R$ belong to the eigenspaces with $\nu=1$, $0$, and $2$, respectively.

\section{Low-energy expansion}
Now let us study the expansion of (4.15) in powers of $k$. Here, we assume that the following limits exist: $V(-\infty) \equiv \lim_{x \to -\infty} V(x)$, $f(-\infty) \equiv \lim_{x \to -\infty} f(x)$, and $f'(-\infty) \equiv \lim_{x \to -\infty} f'(x)$. 
(In the last expression, $f'$ means the derivative of $f$.) These limits may be infinite.

For $k=0$, we can easily solve Eq.~(1.5) and obtain
\begin{equation}
\alpha(x,y;k=0)= \cosh \frac{V(y)-V(x)}{2}, 
\qquad
\beta(x,y;k=0)= \sinh \frac{V(y)-V(x)}{2}.
\end{equation}
Substituting (7.1) into (1.7), (5.13), and (5.12) successively, we find
\begin{equation}
\bar \omega(x,y;W,a;k=0) = W, 
\qquad
\bar A(x,y;W,a;k=0) = a.
\end{equation}
Setting $k=0$ in (5.11), and inserting (7.2), we obtain
\begin{equation}
\mathcal{A}^{-1} g(x,W,a)=\int_{-\infty}^x g(z,W,a)\, d z .
\end{equation}
Similarly, on setting $k=0$, condition (4.12) becomes $\lim_{x \to -\infty} h(x,W,a) = 0$ 
[see also Eq.~(5.10)].  Indeed, the inverse of $\mathcal{A} = \partial/\partial x$ is given by (7.3) if we assume that the domain of $\mathcal{A}$ consists of functions that vanish as $x \to -\infty$.

Let $g$ be a function for which $\mathcal{A}^{-1} g$ makes sense. Since $\mathcal{A}\mathcal{A}^{-1}g=g$, we have the identity
\begin{equation}
g=(\mathcal{A}-2 i k \mathcal{B}) \mathcal{A}^{-1} g + 2 i k \mathcal{B} \mathcal{A}^{-1} g.
\end{equation}
The inverse of $(\mathcal{A}-2 i k \mathcal{B})$ is given by (5.11) if the domain of $(\mathcal{A}-2 i k \mathcal{B})$ is restricted to functions satisfying (4.12). 
This means that $(\mathcal{A}-2 i k \mathcal{B})^{-1} (\mathcal{A}-2 i k \mathcal{B}) \mathcal{A}^{-1} g =\mathcal{A}^{-1} g$ if
\begin{equation}
\lim_{z \to -\infty} \mathcal{U}(x,z) (\mathcal{A}^{-1} g)(z)=0.
\end{equation}
Let us apply $(\mathcal{A}-2 i k \mathcal{B})^{-1}$ to both sides of (7.4).
If (7.5) holds, we have
\begin{equation}
(\mathcal{A}-2 i k \mathcal{B})^{-1}g
=
\mathcal{A}^{-1} g + 2 i k(\mathcal{A}-2 i k \mathcal{B})^{-1} \mathcal{B} \mathcal{A}^{-1} g.
\end{equation}
The expressions $(\mathcal{A}-2 i k \mathcal{B})^{-1}g$ and $\mathcal{A}^{-1} g$ make sense if and only if the integrals in (5.11) and (7.3) are convergent. 
The second term on the right-hand side of (7.6) automatically makes sense if both $(\mathcal{A}-2 i k \mathcal{B})^{-1}g$ and $\mathcal{A}^{-1} g$ make sense.
In summary, (7.6) holds if
\begin{flalign}
\qquad \quad &\text{(i) \ the integrals in (5.11) and (7.3) are both convergent \ \ and} &
\nonumber \\*
\qquad \quad &\text{(ii) \ condition (7.5) holds.} &
\end{flalign}
The $N+1$ times iteration of (7.6) yields
\begin{equation}
(\mathcal{A}-2 i k \mathcal{B})^{-1}g
= 
\sum_{n=0}^N
(2 i k)^n (\mathcal{A}^{-1} \mathcal{B})^n \mathcal{A}^{-1} g 
+ (2 i k)^{N+1}(\mathcal{A}-2 i k \mathcal{B})^{-1} \mathcal{B}(\mathcal{A}^{-1} \mathcal{B})^N \mathcal{A}^{-1} g.
\end{equation}
We assume that $(\mathcal{A}-2 i k \mathcal{B})^{-1}g$ makes sense. Then (7.8) holds if $(\mathcal{A}^{-1} \mathcal{B})^n \mathcal{A}^{-1} g$ makes sense for all $n \leq N$ and if
\begin{equation}
\lim_{z \to -\infty} \mathcal{U}(x,z) [(\mathcal{A}^{-1} \mathcal{B})^n \mathcal{A}^{-1} g](z)=0 \qquad \text{for all $n \leq N$}.
\end{equation}

From (7.8) and (4.15), we obtain
\begin{equation}
\hat R(x,-\infty)
=
\hat r_0
+ i k \hat r_1
+ (i k)^2 \hat r_2
+ (i k)^3 \hat r_3
+ \cdots
+ (i k)^N \hat r_N + \hat \rho_N
\end{equation}
with
\begin{equation}
\hat r_n(x) \equiv (2 \mathcal{A}^{-1} \mathcal{B})^n \mathcal{A}^{-1} \mu^2 f(x),
\end{equation}
\begin{equation}
\hat \rho_N(x,k) \equiv 2 (i k)^{N+1}(\mathcal{A}-2 i k \mathcal{B})^{-1} \mathcal{B}\,\hat r_N(x).
\end{equation}
(We omit to write the dependence on $W$ and $a$.)
From (7.9), we have the condition
\begin{equation}
\lim_{z \to -\infty} \mathcal{U}(x,z) \, \hat r_n(z) = 0.
\end{equation}
Equation (7.10) is valid if 
\begin{flalign}
\qquad \quad &\text{(i) \ the right-hand side of (7.11) makes sense for $n=0,1,\ldots,N$ \  \ and} &
\nonumber \\*
\qquad \quad &\text{(ii) \ condition (7.13) holds for $n=0,1,\ldots,N$.} &
\end{flalign}

Substituting $\mu=a/\cosh \frac{W-V(x)}{2}$ and using (6.4), we can write (7.11) as
\begin{equation}
\hat r_n(x) = a^2 \left[2 (\mathcal{A}^{(2)})^{-1} \mathcal{B}^{(2)}\right]^n (\mathcal{A}^{(2)})^{-1}\frac{f(x)}{\cosh^2 \frac{W-V(x)}{2}},
\end{equation}
where $\mathcal{B}^{(2)}$ is given by (6.5) with $\nu=2$, and
\begin{equation}
(\mathcal{A}^{(2)})^{-1} g(x,W)=\int_{-\infty}^x g(z,W)\,d z.
\end{equation}
Since $f(x)/\cosh^2 \frac{W-V(x)}{2} = (d/dx) \tanh \frac{W-V(x)}{2}$, we obtain $\hat r_0$ from (7.15) as
\begin{equation}
\hat r_0(x) = 
a^2
\left(
\tanh \frac{W-V(x)}{2} - \tanh \frac{W-V(-\infty)}{2}
\right).
\end{equation}
We need to consider three cases, $V(-\infty)\neq \pm \infty$, $V(-\infty)=+\infty$, and $V(-\infty)=- \infty$.  

\subsection{The case $\boldsymbol{V(-\infty)=V_1 \neq \pm \infty}$}
First, we consider the case where $V(-\infty)$ has a finite value $V_1$. 
We can rewrite (7.17) in the form
\begin{subequations}
\begin{equation}
\hat r_0(x) =  \frac{a^2 \sinh \frac{V_1 - V(x)}{2}}{\cosh \frac{W - V_1}{2}\,\cosh \frac{W - V(x)}{2}}.
\end{equation}
Using $\hat r_n= \left[2 (\mathcal{A}^{(2)})^{-1} \mathcal{B}^{(2)}\right]^n \hat r_0$, the coefficients for $n \geq 1$ can be calculated as\cite{note3}
\begin{align}
\hat r_1(x) &=
\frac{a^2}{\cosh^2 \frac{W-V_1}{2}}
\int_{-\infty}^x d z \,\sinh[V_1 - V(z)], &
\\
\hat r_2(x) &=
\frac{a^2}{\cosh^3 \frac{W-V_1}{2}}
\Biggl\{{}
e^{(W+V_1)/2} \int_{-\infty}^x \! \! d z \int_{-\infty}^z \! \! d z' \,e^{-V(z)}
\sinh[V_1 - V(z')]
\nonumber \\*
& \qquad \qquad \qquad \qquad+ e^{-(W+V_1)/2} \int_{-\infty}^x \! \! d z \int_{-\infty}^z \! \! d z' \,e^{V(z)}
\sinh[V_1 - V(z')]
\Biggr\},
\end{align}
\end{subequations}
and so on. It can be easily shown by induction that $\hat r_n$ ($n \geq 1$) has the form
\begin{equation}
\hat r_n(x)
=
\frac{a^2}{\cosh^{n+1} \frac{W-V_1}{2}}
\sum_{j=0}^{n-1} \exp \! \left[\left(j-\tfrac{n-1}{2}\right) W\right] C_j(x),
\end{equation}
where $C_j(x)$ is a function that vanishes as $x \to -\infty$, provided that $\hat r_n$ makes sense.
Let $F^{(-)}_n$ denote the set of real-valued functions $g(x)$ such that
\begin{equation}
\int_{-\infty}^{x_0} (1+ \vert x \vert^n) \vert g(x) \vert \, d x < \infty 
\quad \mbox{for any finite $x_0$}.
\end{equation}
It is not difficult to show that $\hat r_n$ makes sense if $V-V_1 \in F^{(-)}_{n-1}$ (see Appendix~C of Ref.~\onlinecite{low}). 
Since $F_{N-1}^{(-)} \subset F_{n-1}^{(-)}$ for $n < N$, condition (i) of (7.14) is satisfied if $V-V_1 \in F^{(-)}_{N-1}$. 

We also need to check condition (ii) of (7.14). 
From (5.10), (5.12), and (7.18a) we obtain
\begin{equation}
\mathcal{U}(x,z)\, \hat r_0(z)
=\frac{- \hat T^2(x,z) \tanh{\frac{V(z) - V_1}{2}}}{1 + \hat L(x,z)  \tanh{\frac{V(z) - V_1}{2}}}.
\end{equation}
Similarly, from (5.10), (5.12), and (7.19) we have, for $n \geq 1$,
\begin{equation}
\mathcal{U}(x,z)\, \hat r_n(z)
=\frac{\hat T^2(x,z)}{\cosh^{n+1} \frac{V(z) - V_1}{2}}
  \sum_{j=0}^{n-1}\exp \! \left[\left(j-\tfrac{n-1}{2}\right) V(z)\right] B_j(x,z)\, C_j(z),
\end{equation}
where
\begin{equation}
B_j(x,z) \equiv 
\bigl[1 + \hat L(x,z)\bigr]^j \bigl[1 - \hat L(x,z)\bigr]^{n-j-1}
\left[1 + \hat L(x,z) \tanh{\tfrac{V(z) - V_1}{2}}\right]^{-(n+1)}.
\end{equation}
The asymptotic behavior of $\hat T(x,z)$ and $\hat L(x,z)$ as $z \to -\infty$ is shown in Appendix~C. 
As we are assuming that $f(-\infty)$ exists, $V(-\infty) \neq \pm \infty$ implies $f(-\infty)=0$. From (C6) of Appendix~C, we can see that $\lim_{z \to -\infty} \vert B_j(x,z) \vert < \infty$. Since $\tanh{\frac{V(z) - V_1}{2}} \to 0$ and $C_j(z) \to 0$ as $z \to -\infty$, the right-hand sides of (7.21) and (7.22) vanish in this limit.  This means that (7.13) holds as long as $\hat r_n$ makes sense.  Both conditions (i) and (ii) of (7.14) hold, and hence, (7.10) is valid, if $V-V_1 \in F^{(-)}_{N-1}$.

\subsection{The case $\boldsymbol{V(-\infty)=+\infty}$}
Next, we study the case $V(-\infty)=+\infty$. Now (7.17) becomes
\begin{subequations}
\begin{equation}
\hat r_0(x) = 2 a^2 \left(1 + e^{V(x) - W}\right)^{-1}.
\end{equation}
Using $\hat r_n= \left[2 (\mathcal{A}^{(2)})^{-1} \mathcal{B}^{(2)}\right]^n \hat r_0$, we can calculate
\begin{align}
&\hat r_1(x)= 2 a^2 e^W\int_{-\infty}^x d z\, e^{-V(z)},
\qquad
\hat r_2(x)= 4 a^2 e^{2 W} \int_{-\infty}^x  \! \! d z \int_{-\infty}^z  \! \!  d z'\,e^{-V(z) - V(z')},
\end{align}
\end{subequations}
and so on. For general $n \geq 1$, the expression for $\hat r_n$ can be explicitly obtained as\cite{low}
\begin{equation}
\hat r_n(x)
= a^2 \sum_{\sigma_2=\pm 1} \sum_{\sigma_3=\pm 1} \cdots \sum_{\sigma_n =\pm1}
C_{+1,\sigma_2,\ldots,\sigma_n} 
\exp\Bigl[
\Bigl(
1 + 
\sum_{j=2}^n \sigma_j
\Bigr)
W
\Bigr]
I_{+1,\sigma_2,\ldots,\sigma_n}(x),
\end{equation}
where
\begin{equation}
I_{\sigma_1,\sigma_2,\ldots,\sigma_n} (x) 
\equiv \int_{-\infty}^x \! \! d z_n \int_{-\infty}^{z_n} \! \! d z_{n-1} \cdots \int_{-\infty}^{z_3} \! \! d z_2 \int_{-\infty}^{z_2} \! \! d z_1 \, 
\exp\Bigl[-\sum_{j=1}^n \sigma_j V(z_j)\Bigr],
\end{equation}
\begin{equation}
C_{\sigma_1,\sigma_2,\ldots,\sigma_n}
\equiv 2 \prod_{j=1}^n \Bigl[\Bigl( \sum_{i=1}^j \sigma_i \Bigr)\sigma_j\Bigr].
\end{equation}
It can be seen from (7.27) that $C_{+1,\sigma_2,\ldots, \sigma_n} = 0$ unless
\begin{equation}
1 + \sum_{i=2}^j \sigma_i >0
\qquad \mbox{for $j=2,3,\ldots,n$}.
\end{equation}
So, only the terms with $\sigma_2,\ldots,\sigma_n$ satisfying (7.28) are present in (7.25). [In particular, only $\sigma_2=+1$ survives in (7.25).] It can be shown\cite{low} that $\vert I_{+1,\sigma_2,\ldots,\sigma_n} \vert < \infty$ for all $\sigma_2,\ldots,\sigma_n$ satisfying (7.28) if $e^{-V} \in F^{(-)}_{n-1}$. Therefore, $\hat r_n$ makes sense if $e^{-V} \in F^{(-)}_{n-1}$, and condition (i) of (7.14) is satisfied if $e^{-V} \in F^{(-)}_{N-1}$.

We proceed to check condition (ii) of (7.14). Using (5.10) with (7.24a) and (7.25), we obtain\begin{equation}
\mathcal{U}(x,z) \, \hat r_0(z)
=\frac{\hat T^2(x,z)}{1 - \hat L(x,z)},
\end{equation}
\begin{equation}
\mathcal{U}(x,z) \, \hat r_n(z)
=
\hat T^2(x,z)
\sum_{\sigma_2=\pm 1} \sum_{\sigma_3=\pm 1} \cdots \sum_{\sigma_n =\pm1}
C_{+1,\sigma_2,\ldots,\sigma_n}
B_{\sigma_2, \ldots, \sigma_n}(x,z)\, D_{\sigma_2, \ldots, \sigma_n}(z),
\end{equation}
where
\begin{align}
&B_{\sigma_2, \ldots, \sigma_n}(x,z)
\equiv
\left[ 1 + \hat L(x,z) \right]^{\sum_{j=2}^n \sigma_j}
\left[ 1 - \hat L(x,z) \right]^{-2 -\sum_{j=2}^n \sigma_j},
\\
&D_{\sigma_2, \ldots, \sigma_n}(z)
\equiv
\exp\Bigl[
\Bigl(
1 + \sum_{j=2}^n \sigma_j
\Bigr) V(z)
\Bigr]
I_{+1,\sigma_2,\ldots,\sigma_n}(z).
\end{align}

 When $V(-\infty)=+\infty$, there are three possible cases for $f(-\infty)$, namely, $f(-\infty)=+\infty$, $f(-\infty)=c>0$, and $f(-\infty)=0$.
Let us first assume that $f(-\infty) \neq 0$.
From (C1) and (C3) of Appendix~C, we know that if $f(-\infty)=+\infty$ or $f(-\infty)=c > 0$,
\begin{equation}
\hat T(x,z) \to 0, \qquad \hat L(x,z) \to \mbox{constant}\neq 1 
\qquad \mbox{as $z \to -\infty$}.
\end{equation}
(Here, we are assuming that $ c > \vert k \vert$, as we are now interested in the low-energy region.)
In this case, we can easily show that $\lim_{z \to -\infty} \vert D_{\sigma_2, \ldots, \sigma_n}(z) \vert < \infty$ for all $\sigma_2,\ldots,\sigma_n$ satisfying (7.28). On account of (7.28), we have $\sum_{j=2}^n \sigma_j \geq 0$. From this and (7.33), it follows that
$
\lim_{z \to -\infty} \vert B_{\sigma_2, \ldots, \sigma_n}(x,z) \vert <\infty.
$
Now it is obvious from (7.33) that the right-hand sides of (7.29) and (7.30) vanish in the limit $z \to -\infty$. 
Thus, (7.13) holds if $\hat r_n$ makes sense.

The situation is different for the case $f(-\infty)=0$. 
When $V(-\infty)=+\infty$ and $f(-\infty)=0$, condition (7.13) does not hold for $\mathrm{Im}\,k=0$. 
As can be seen from (C4), in this case, $\hat T(x,z)$ oscillates as $z \to -\infty$ for $\mathrm{Im}\,k=0$, so neither (7.29) nor (7.30) becomes zero in the limit $z \to -\infty$.
However, this does not mean that (7.10) is not correct for $\mathrm{Im}\,k =0$. 
Corresponding to (1.9), the definition of $\hat R$ for real $k$ is $\hat R(k)= \lim_{\epsilon \downarrow 0} \hat R(k + i \epsilon)$. If conditions (7.14) hold for $\mathrm{Im}\,k>0$, Eq.~(7.10) is valid even for $\mathrm{Im}\,k = 0$  as long as remainder term (7.12) is interpreted as  $\lim_{\epsilon \downarrow 0}\hat \rho_N(x, k + i \epsilon)$.

Let us assume $\mathrm{Im}\,k>0$ to consider the case where $V(-\infty)=+\infty$ and $f(-\infty)=0$.  It can be seen from (C4) that $\hat T(x,z)$ approaches zero exponentially as $z \to -\infty$. 
Meanwhile, $D_{\sigma_2, \ldots, \sigma_n}(z)$ defined by (7.32) becomes infinite in this limit; it grows at most like $\vert z \vert^n$ as $z \to -\infty$ if $\hat r_n$  makes sense. Since $\hat T^2(x,z)$ falls off exponentially and $\vert B_{\sigma_2, \ldots, \sigma_n}(x,z) \vert$ remains finite, both (7.29) and (7.30) vanish in the limit  $z \to -\infty$.

Thus, in the case $V(-\infty) = +\infty$, Eq.~(7.13) holds (at least for $\mathrm{Im}\,k>0$)  if $e^{-V} \in F^{(-)}_{n-1}$, irrespective of whether $f(-\infty)\neq 0$ or $f(-\infty)=0$. If $e^{-V} \in F^{(-)}_{N-1}$, both conditions (i) and (ii) of (7.14) hold for $\mathrm{Im}\,k>0$, and hence, (7.10) is valid for $\mathrm{Im}\,k \geq 0$.

\subsection{The case $\boldsymbol{V(-\infty)=-\infty}$}
The case $V(-\infty) = -\infty$ is similar to the case $V(-\infty)= + \infty$. 
In this case we have
\begin{align}
\hat r_n=- a^2 \sum_{\sigma_2=\pm 1} \sum_{\sigma_3=\pm 1} \cdots \sum_{\sigma_n =\pm1}
C_{-1,\sigma_2,\ldots,\sigma_n} 
\exp\Bigl[
\Bigl(
-1 +
\sum_{j=2}^n \sigma_j
\Bigr)
W
\Bigr]
I_{-1,\sigma_2,\ldots,\sigma_n}(x).
\end{align}
In the same way as in the previous case, we can show that (7.10) is valid if $e^{V} \in F^{(-)}_{N-1}$.

\bigskip
\bigskip
\bigskip
Thus, we have found that condition~(ii) of (7.14) is always satisfied as long as condition~(i) holds.
Equation (7.10) is valid if the coefficients $\hat r_0, \hat r_1, \ldots, \hat r_N$ make sense.

For (7.10) to be an asymptotic expansion, the remainder term $\hat \rho_N$ must satisfy
\begin{equation}
\lim_{k \to 0} \hat \rho_N(x,k)/k^N=0.
\end{equation}
Using (5.11), the expression for remainder term (7.12) can be written as
\begin{equation}
\hat \rho_N = 2 (2 i k)^{N+1} \int_{-\infty}^x dz\, 
\frac{\hat T^2(x,z;W)}{1 - \hat L^2(x,z;W)}\,K_N(z,\bar \omega(x,z;W)),
\end{equation}
where $K_N(x,W) \equiv (1/a^2)\,\mathcal{B}\, \hat r_N$.
We can investigate the validity of (7.35) by using this integral form. 
This was essentially done in Refs.~\onlinecite{analysis, low}.  It can be shown that (7.35) is satisfied if $V-V_1 \in F^{(-)}_{N-1}$ or $e^{\mp V} \in F^{(-)}_{N-1}$.
(See Appendix~E of Ref.~\onlinecite{low} for details.)

\section{Asymptotically periodic potentials}
In Sec.~VII, it was assumed that $V(-\infty)$ exists. An important class of potentials that does not fall into this category is asymptotically periodic potentials,\cite{asymptotic} i.e., potentials that tend to a periodic function as $x \to -\infty$. [Here, the term ^^ ^^ potential" refers to the Fokker-Planck potential $V(x)$. The corresponding Schr\"odinger potential $V_\mathrm{S}(x)$ is also asymptotically periodic under the following assumptions.]
In this section, we consider such potentials.
 We assume that $V$ and $f$ have the forms
\begin{equation}
V(x)= V_\mathrm{p}(x) + V_\Delta (x),
\qquad
f(x)= f_\mathrm{p}(x) + f_\Delta (x),
\end{equation}
where $V_\mathrm{p}$ and $f_\mathrm{p}$ are periodic functions with period $L$, while
 $V_\Delta$ and $f_\Delta$ vanish as $x \to -\infty$,
\begin{equation}
V_\mathrm{p}(x + L) = V_\mathrm{P}(x),
\quad 
f_\mathrm{p}(x + L) = f_\mathrm{p}(x),
\quad 
\lim_{x \to -\infty} V_\Delta(x) =0, 
\quad 
\lim_{x \to -\infty} f_\Delta(x) =0. 
\end{equation}
We also assume that the derivative of $f_\Delta(x)$ vanishes as $x \to -\infty$.

For such $V(x)$, expression (7.17) does not make sense. This is because (7.3) does not give the inverse of $\mathcal{A}$ when the domain of $\mathcal{A}$ consists of functions that do not vanish as $x \to -\infty$. 
We need to find the appropriate inverse of $\mathcal{A}$ for asymptotically periodic functions.

Let us consider Eq.~(4.3) with $k=0$,
\begin{equation}
\mathcal{A} h(x,W,a) = \frac{\partial}{\partial x} h(x,W,a) = g(x,W,a),
\end{equation}
assuming that both $h$ and $g$ are asymptotically periodic functions of $x$,
\begin{subequations}
\begin{equation}
h = h_\mathrm{p} + h_\Delta,  \qquad g = g_\mathrm{p} + g_\Delta,
\end{equation}
\begin{equation}
h_\mathrm{p}(x + L) = h_\mathrm{p}(x), \quad
g_\mathrm{p}(x + L) = g_\mathrm{p}(x), \quad
\lim_{x \to -\infty} h_\Delta(x) =0, \quad
\lim_{x \to -\infty} g_\Delta(x) =0.
\end{equation}
\end{subequations}
We also assume that $h$ and $g$ are analytic functions of $W$ in the region $\vert \mathrm{Im}\,W \vert < \pi/2$. (See the comment at the end of Sec.~V.)
Equations (8.3) and (8.4) imply $g_\Delta = \partial h_\Delta/\partial x$ and $g_\mathrm{p} = \partial h_\mathrm{p}/\partial x$.
Hence, it follows that
\begin{equation}
\int_{x-L}^x g_\mathrm{p}(z, W, a) \, d z = 0
\end{equation}
for any $x$.
The relation $g_\Delta = \partial h_\Delta/\partial x$ can be inverted as
\begin{equation}
h_\Delta(x,W,a) = \int_{-\infty}^x g_\Delta(z,W,a) \, d z.
\end{equation}
On the other hand, $g_\mathrm{p} = \partial h_\mathrm{p}/\partial x$ cannot be uniquely inverted.
All we have is
\begin{equation}
h_\mathrm{p}(x,W,a) = \int_{x_0}^x g_\mathrm{p}(z,W,a) \, d z + C_{x_0}(W,a),
\end{equation}
where $x_0$ is an arbitrarily fixed number, and $C_{x_0}(W,a)$ is an $x$-independent function which is undetermined at this moment.

When $g$ is given, the function $h$ satisfying (8.3) is not uniquely determined by conditions (8.4), as $C_{x_0}(W,a)$ is undetermined. However, assuming that we have a further condition to determine $C_{x_0}(W,a)$ (such a condition will be shortly shown), the inverse of $\mathcal{A}$ can be written as
$\mathcal{A}^{-1} g = h_\mathrm{p} + h_\Delta$.
We denote the right-hand side of (8.7) by $\mathcal{A}_\mathrm{p}^{-1} g_\mathrm{p}(x,W,a)$.
Then,
\begin{equation}
\mathcal{A}^{-1} g(x,W,a)= \int_{-\infty}^x g_\Delta(z,W,a) \, d z + \mathcal{A}_\mathrm{p}^{-1} g_\mathrm{p}(x,W,a).
\end{equation}
It should be noted that $\mathcal{A}^{-1} g$ does not make sense unless $g_\mathrm{p}$ satisfies (8.5).

Using this $\mathcal{A}^{-1}$, we shall derive the expansion of $(\mathcal{A}  - 2 i k \mathcal{B})^{-1}$ as we did in Sec.~VII.
The starting point is Eq.~(7.6). 
Although $C_{x_0}(W,a)$ in (8.7) has not yet been determined, Eq.~(7.6) holds for any $C_{x_0}$ as long as condition (7.5) is satisfied.
However, for (7.6) to be useful for low energy expansion, it is necessary that
\begin{equation}
\lim_{k \to 0} 2 i k (\mathcal{A} - 2 i k \mathcal{B})^{-1} \mathcal{B} \mathcal{A}^{-1}  g = 0
\end{equation}
so that (7.6) may yield
$\lim_{k \to 0} (\mathcal{A} - 2 i k \mathcal{B})^{-1} g = \mathcal{A}^{-1} g$. 
We also require that
\begin{equation}
\lim_{k \to 0} (\mathcal{A} - 2 i k \mathcal{B})^{-1} \mathcal{B} \mathcal{A}^{-1}  g = \mathcal{A}^{-1} \mathcal{B} \mathcal{A}^{-1}g.
\end{equation}
Then (8.9) holds if $\mathcal{A}^{-1} \mathcal{B} \mathcal{A}^{-1} g$ makes sense.
In order that $\mathcal{A}^{-1} \mathcal{B} \mathcal{A}^{-1} g$ may make sense, the periodic part of  
$\mathcal{B} \mathcal{A}^{-1} g$ must satisfy the condition corresponding to (8.5). 
The periodic part of $\mathcal{B} \mathcal{A}^{-1} g$ is $\mathcal{B}_\mathrm{p} \mathcal{A}_\mathrm{p}^{-1} g_\mathrm{p}$ with $\mathcal{B}_\mathrm{p}$ defined by
\begin{equation}
\mathcal{B}_\mathrm{p} \equiv \sinh [W - V_\mathrm{p}(x)] \frac{\partial}{\partial W} + \frac{1}{2} \cosh [W - V_\mathrm{p} (x)] \, a \frac{\partial}{\partial a}.
\end{equation} 
Putting $\mathcal{B}_\mathrm{p} \mathcal{A}_\mathrm{p}^{-1} g_\mathrm{p}$ in place of $g_\mathrm{p}$ in (8.5), we have
\begin{equation}
\int_{x-L}^x (\mathcal{B}_\mathrm{p} \mathcal{A}_\mathrm{p}^{-1} g_\mathrm{p})(z,W,a) \, d z = 0.
\end{equation}
The function $C_{x_0}(W,a)$ is determined by condition (8.12).

It is easier to work in the eigenspace of $\mathcal N$ and deal with $\mathcal{A}^{(\nu)}$ and $\mathcal{B}^{(\nu)}$ defined by (6.5).  We assume that $g(x,W) = g_\mathrm{p}(x,W) + g_\Delta(x,W)$ as in (8.4). Corresponding to (8.8), we have
\begin{align}
&(\mathcal{A}^{(\nu)})^{-1} g(x,W) 
= \int_{-\infty}^x g_\Delta(z,W) \, d z
+ (\mathcal{A}^{(\nu)}_\mathrm{p})^{-1}g_\mathrm{p}(x,W) ,
\\
&(\mathcal{A}^{(\nu)}_\mathrm{p})^{-1}g_\mathrm{p}(x,W) 
 \equiv \int_{x_0}^x g_\mathrm{p}(z,W) \, d z + C_{x_0}^{(\nu)}(W).
\end{align}
The function $C_{x_0}^{(\nu)}(W)$ in (8.14) is determined from the condition corresponding to (8.12),
 \begin{equation}
\int_{x-L}^x [\mathcal{B}_\mathrm{p}^{(\nu)} (\mathcal{A}_\mathrm{p}^{(\nu)})^{-1} g_\mathrm{p}](z,W) \, d z = 0,
\end{equation}
where
\begin{equation}
\mathcal{B}_\mathrm{p}^{(\nu)} \equiv 
\sinh [W - V_\mathrm{p}(x)] 
\frac{\partial}{\partial W}
+ \frac{\nu}{2} \cosh[ W - V_\mathrm{p}(x) ].
\end{equation}
Note that $(\mathcal{A}^{(\nu)})^{-1}$ explicitly depends on $\nu$ although the expression of $\mathcal{A}^{(\nu)}$  in (6.5) appears to be $\nu$-independent.
This is because the appropriate domain of $\mathcal{A}^{(\nu)}$ depends on~$\nu$.

For the analysis of $\hat R$, we need only to deal with $\nu=2$. 
Let us determine $C_{x_0}^{(2)}(W)$ from condition (8.15).
For $\nu = 2$, substituting (8.14) and (8.16) into (8.15) gives
\begin{equation}
\int_{x-L}^x d z
\,\frac{\partial}{\partial W}
\left\{
\sinh[W - V_\mathrm{p}(z)]
\left[
\int_{x_0}^z d z' g_\mathrm{p}(z',W) + C_{x_0}^{(2)}(W)
\right]
\right\}
= 0.
\end{equation}
We define the constants $L_0$ and $V_0$ by
\begin{equation}
L_0 \equiv \sqrt{
\textstyle \left(\int_{x-L}^x e^{V_\mathrm{p}(z)}\, d z\right)
\left(\int_{x-L}^x e^{- V_\mathrm{p}(z)}\, d z\right)
},
\qquad
V_0 \equiv \frac{1}{2}\log
\frac{\int_{x-L}^x e^{V_\mathrm{p}(z)}\, d z}{\int_{x-L}^x e^{-V_\mathrm{p}(z)}\, d z},
\end{equation}
so that $\int_{x-L}^x e^{\pm V_\mathrm{p}(z)} d z = L_0 \,e^{\pm V_0}$.
Integrating (8.17) with respect to $W$, we obtain
\begin{equation}
C_{x_0}^{(2)}(W) =\frac{-1}{L_0 \sinh (W - V_0)}
\left\{
\int_{x-L}^x d z \int_{x_0}^z d z' \sinh [W - V_\mathrm{p}(z)] g_\mathrm{p}(z',W) + C
\right\},
\end{equation}
where $C$ is a constant of integration.
By our assumption, $C_{x_0}^{(2)}(W)$ should be analytic in the strip $\vert \mathrm{Im}\,W \vert < \pi/2$, so the right-hand side of (8.19) should not be singular at $W=V_0$. From this condition, the constant $C$ in (8.19) is determined as
\begin{equation}
C= - \int_{x-L}^x d z \int_{x_0}^z d z' \sinh [V_0 - V_\mathrm{p}(z)] g_\mathrm{p}(z',V_0).
\end{equation}
Since $x_0$ is arbitrary, we may let $x_0= x-L$. 
Substituting (8.19) into (8.14), letting $x_0 = x-L$, and using $\int_{x-L}^x g_\mathrm{p}(z) \,d z=0$, we obtain 
\begin{multline}
(\mathcal{A}_\mathrm{p}^{(2)})^{-1} g_\mathrm{p}(x,W)
=\frac{1}{L_0 \sinh (W - V_0)}
\\*
\times
\int_{x-L}^x d z \int_{x-L}^z d z' \, 
\bigl\{
 \sinh [V_\mathrm{p}(z) - W] g_\mathrm{p}(z',W)
- \sinh [V_\mathrm{p}(z) - V_0] g_\mathrm{p}(z',V_0)
\bigr\}.
\end{multline}
Thus, with $(\mathcal{A}_\mathrm{p}^{(2)})^{-1}$ given by (8.21), the inverse of $\mathcal{A}^{(2)}$ cam be written as
\begin{equation}
(\mathcal{A}^{(2)})^{-1} g(x,W)
=(\mathcal{A}_\mathrm{p}^{(2)})^{-1} g_\mathrm{p}(x,W)
+  \int_{-\infty}^x g_\Delta(z,W)\, d z.
\end{equation}
The same expression as (8.21) was derived in Ref.~\onlinecite{periodic} by a different method. Although the method of Ref.~\onlinecite{periodic} is meaningful for its own sake and for its applications, expression (8.21) itself can be derived more simply from condition~(8.15) as we have seen above.

We can calculate the coefficients $\hat r_n$ of (7.10) by using (8.22), instead of (7.16), in (7.15).
The calculation is the same as that for the expansion of $\bar R_r$ studied in Ref.~\onlinecite{asymptotic}.
We obtain 
\begin{subequations}
\begin{align}
\hat r_0(x) &= a^2 \left(\tanh \frac{W-V(x)}{2} - \tanh \frac{W-V_0}{2} \right)
=
\frac{a^2 \sinh \frac{V_0 - V(x)}{2}}{\cosh \frac{W - V_0}{2}\,\cosh \frac{W - V(x)}{2}},
\\
\hat r_1(x) &=\frac{a^2}{\cosh^2 \frac{W-V_0}{2}} 
\Biggl({}\frac{1}{2 L_0}
\int_{x-L}^x d z \int _{x-L}^z d z' \sinh[V_\mathrm{p}(z') - V_\mathrm{p}(z)]
\nonumber \\*
 & \qquad \qquad \qquad \qquad+\int_{-\infty}^x d z\left\{\sinh [V_0 - V(z)] -\sinh [V_0 - V_\mathrm{p}(z)] \right\}
\Biggr),
\end{align}
\end{subequations}
and so on.
It can be shown\cite{asymptotic} that $\hat r_n$ makes sense if $V_\Delta \in F_{n-1}^{(-)}$. So, condition (i) of (7.14) is satisfied if $V_\Delta \in F_{N-1}^{(-)}$.
When $\hat r_n$ ($n \geq 1$) makes sense, it has the form
\begin{equation}
\hat r_n(x)
= a^2
\sum_{j=1}^m H_j(W) \left[ C_j^\mathrm{p} (x) + C_j^\Delta (x)\right],
\end{equation}
where $m$ is a finite number, each $C_j^\mathrm{p}(x)$ is a periodic function, each $C_j^\Delta(x)$ is a function that vanishes as $x \to -\infty$, and each $H_j(W)$ is an analytic function of $W$ in the strip $\vert \mathrm{Im}\,W \vert < \pi/2$. 
[Note that (8.23b) has the form of (8.24) with $m=1$.]

Let us check condition (ii) of (7.14). 
From (8.23a), (8.24), and (5.10), we obtain
\begin{equation}
\mathcal{U}(x,z)\, \hat r_0(z)
=\frac{- \hat T^2(x,z) \tanh{\frac{V(z) - V_0}{2}}}{1 + \hat L(x,z)  \tanh{\frac{V(z) - V_0}{2}}},
\end{equation}
and, for $n \geq 1$,
\begin{equation}
\mathcal{U}(x,z)\, \hat r_n(z)
=
\hat T^2(x,z;W,a)
\sum_{j=1}^m
\frac{H_j(\bar \omega(x,z;W))}{1 - \hat L^2(x,z;W)}
\left[ C_j^\mathrm{p} (z) + C_j^\Delta (z)\right].
\end{equation}
Unlike (7.21) and (7.22), the right-hand sides of (8.25) and (8.26) do not tend to zero as $z \to -\infty$ if $\mathrm{Im}\,k=0$. 
In this case, we need to assume $\mathrm{Im}\,k >0$. As mentioned in Sec.~VII, this is enough for the validity of (7.10) in the closed half plane $\mathrm{Im}\,k \geq 0$.

Let $U_\mathrm{p}(x,y;k)$ be the $2\times 2$ matrix defined as the solution of (1.5)  with $f$ replaced by $f_\mathrm{p}$, satisfying the boundary condition $U_\mathrm{p}(y,y;k)=I$. Let $\lambda$ be the eigenvalue of $U_\mathrm{p}(x, x-L;k)$ such that $\vert \lambda \vert > 1$. (Such $\lambda$ uniquely exists for $\mathrm{Im}\,k>0$, since the two eigenvalues of $U_\mathrm{p}$ are reciprocal of each other. This $\lambda$ is independent of $x$. See Ref.~\onlinecite{periodic} for details.) The asymptotic form of $\hat T$ for $\mathrm{Im}\,k > 0$ can be written in terms of this eigenvalue $\lambda$,
\begin{equation}
\hat T(x,z;W,a) \sim a \lambda^{z/L} C(x,z;W)  \qquad \mbox{as $z \to -\infty$},
\end{equation}
where $C(x,z;W)$ is a periodic function of $z$.
This implies that $\hat T(x,z) \to 0$ as $z \to -\infty$, since $\vert \lambda \vert > 1$. 
The asymptotic form of $\hat L$ is
\begin{equation}
\hat L(x,z;W) \sim R_l^\mathrm{p}(\infty,z)
=\frac{\beta_\mathrm{p}(z+L,z;-k)}{\lambda^{-1} - \alpha_\mathrm{p}(z+L,z;k)}
\qquad
\mbox{as $z \to -\infty$},
\end{equation}
where $\alpha_\mathrm{p}$ and $\beta_\mathrm{p}$ are the elements of the matrix $U_\mathrm{p}$ defined in the same way as in (1.6), and $R_l^\mathrm{p} \equiv - \beta_\mathrm{p}(-k)/\alpha_\mathrm{p}(k)$ as in (1.7). [See Eq.~(D.7) of Ref.~\onlinecite{periodic}.]
This $R_l^\mathrm{p}(\infty, z)$ is a periodic function of $z$, and $\vert R_l^\mathrm{p}(\infty,z) \vert < 1$ for $\mathrm{Im}\,k >0$. 

From (8.28) and (5.12), we can see that both $\hat L(x,z)$ and $\bar \omega(x,z;W)$ tend to periodic functions as $z$ approaches $-\infty$.
Hence, each term in the sum on the right-hand side of (8.26) tends to a periodic function. These terms are always finite.
 Since $\hat T(x,z) \to 0$ as $z \to -\infty$, the right-hand sides of (8.25) and (8.26) vanish in this limit. Thus, (7.13) holds if $\hat r_n$ makes sense.
In this case, too, condition (ii) of (7.14) is satisfied if condition (i) holds.

It can also be shown\cite{asymptotic} that (7.35) holds if $V_\Delta \in F_{N-1}^{(-)}$. Therefore, for asymptotically periodic potentials, (7.10) is valid as an asymptotic expansion to order $k^N$ if $V_\Delta \in F_{N-1}^{(-)}$.

\section{High-energy expansion}

In this section, we study the expansion of (4.15) for large $\vert k \vert$.
As $\vert k \vert$ becomes large, it is expected that $(\mathcal{A} - 2 i k \mathcal{B})^{-1}$ will approach $-(2 i k)^{-1} \mathcal{B}^{-1}$ in some sense. However, the inverse of $\mathcal{B}$ does not exist if we let the domain of $\mathcal{B}$ include all analytic functions of $\xi$ and $\mu$.
Instead, we confine ourselves to the eigenspace of $\mathcal{N}$ with eigenvalue $\nu>0$, in which $\mathcal{B}$ reduces to $\mathcal{B}_{(\nu)}$ defined by (6.9).
We let the domain of $\mathcal{B}_{(\nu)}$ consist of functions of $\xi$ which are analytic in the region $\vert \xi \vert < 1$. [See the comment above Eq.~(3.3).]
Then, the inverse of $\mathcal{B}_{(\nu)}$ exists as shown below.
Suppose that $\mathcal{B}_{(\nu)} h= g$. With (6.9), this equation reads 
\begin{equation}
\left(\xi \frac{\partial}{\partial \xi} + \frac{\nu}{2} \right) h(\xi)
=
\xi^{1 - (\nu/2)}\frac{\partial}{\partial \xi}
\left[\xi^{\nu/2} h(\xi) \right]
= g(\xi).
\end{equation}
Hence, $h(\xi)=\xi^{-(\nu/2)} \int_{\xi_0}^\xi \xi^{-1 + (\nu/2)} g(\xi)\, d \xi$, 
where $\xi_0$ is a constant. 
As $h(\xi)$ is analytic in $\vert \xi \vert <1$, this  $\xi_0$ must be zero. [Otherwise, $h(\xi)$ becomes singular at $\xi=0$.] Thus, the inverse of $\mathcal{B}_{(\nu)}$ is
\begin{equation}
\mathcal{B}_{(\nu)}^{-1}\,g(\xi) = \xi^{-(\nu/2)} \int_0^\xi \xi^{-1 + (\nu/2)} g(\xi)\, d \xi.
\end{equation}

Here, we take $\{x,\xi,\mu\}$ as the independent variables rather than $\{x, W, a\}$. Eigenfunctions of $\mathcal{N}$ have the form $\mu^\nu g(x,\xi)$. 
Since $\mathcal{B}\, \mu^\nu \mathcal{B}_{(\nu)}^{-1}\, g = \mu^\nu\mathcal{B}_{(\nu)}^{\phantom{-1}} \mathcal{B}_{(\nu)}^{-1}\, g =  \mu^\nu g$, we can write
\begin{equation}
\mu^\nu g(x,\xi)
 =- \frac{1}{2 i k}
 \left[
 (\mathcal{A} - 2 i k \mathcal{B}) - \mathcal{A}
\right]
\mu^\nu \mathcal{B}_{(\nu)}^{-1} \, g(x,\xi).
\end{equation}
We have $(\mathcal{A} - 2 i k \mathcal{B})^{-1}(\mathcal{A} - 2 i k \mathcal{B}) \mu^\nu \mathcal{B}_{(\nu)}^{-1} \, g = \mu^\nu \mathcal{B}_{(\nu)}^{-1} \, g$
if
\begin{equation}
\lim_{z \to -\infty} \mathcal{U}(x,z) \mu^\nu (\mathcal{B}_{(\nu)}^{-1} \, g)(z,\xi)
=0.
\end{equation}
We assume that $(\mathcal{A} - 2 i k \mathcal{B})^{-1} \mu^\nu g$ makes sense.
Letting $(\mathcal{A} - 2 i k \mathcal{B}) ^{-1}$ act on (9.3), we obtain
\begin{align}
(\mathcal{A} - 2 i k \mathcal{B})^{-1} \mu^\nu g
= - \frac{\mu^\nu}{2 i k} \mathcal{B}_{(\nu)}^{-1}\, g
+ \frac{1}{2 i k} (\mathcal{A} - 2 i k \mathcal{B})^{-1}
\mu^\nu \mathcal{A}_{(\nu)}^{\phantom{-1}} \mathcal{B}_{(\nu)}^{-1} \, g
\end{align}
provided that (9.4) holds.
[We have also used (6.8).]
The $N$ times iteration of (9.5) gives
\begin{align}
(\mathcal{A} - 2 i k \mathcal{B})^{-1} \mu^\nu g
= &{} - \mu^\nu 
\sum_{n=0}^{N-1}
\frac{1}{(2 i k)^{n+1}}
\bigl( \mathcal{B}_{(\nu)}^{-1} \mathcal{A}_{(\nu)}^{\phantom{-1}} \bigr)^n \mathcal{B}_{(\nu)}^{-1} \, g 
\nonumber \\*
& + \frac{1}{(2 i k)^N} 
(\mathcal{A} - 2 i k \mathcal{B})^{-1} \mathcal{A}\mu^\nu  
\bigl(\mathcal{B}_{(\nu)}^{-1} \mathcal{A}_{(\nu)}^{\phantom{-1}} \bigr)^{N-1} \mathcal{B}_{(\nu)}^{-1} \, g.
\end{align}
Equation (9.6) holds if $\bigl( \mathcal{B}_{(\nu)}^{-1} \mathcal{A}_{(\nu)}^{\phantom{-1}} \bigr)^n \mathcal{B}_{(\nu)}^{-1} \, g$ makes sense for all $n \leq N-1$ and if
\begin{equation}
\lim_{z \to -\infty} \mathcal{U}(x,z) \mu^\nu
\left[
\bigl(\mathcal{B}_{(\nu)}^{-1} \mathcal{A}_{(\nu)}^{\phantom{-1}} \bigr)^n \mathcal{B}_{(\nu)}^{-1} \,
 g\right] \! \!(z,\xi)=0 
\qquad \text{for all $n \leq N-1$}.
\end{equation}

Using (9.6) in (4.15) yields
\begin{equation}
\hat R(x,-\infty)
= \frac{1}{2 i k} \hat c_1 + \frac{1}{(2 i k)^2} \hat c_2 +  \frac{1}{(2 i k)^3} \hat c_3
+ \cdots + \frac{1}{(2 i k)^N} \hat c_N + \hat \delta_N
\end{equation}
with
\begin{equation}
\hat c_n(x) \equiv - \mu^2 \bigl(\mathcal{B}_{(2)}^{-1} \mathcal{A}_{(2)}^{\phantom{-1}} \bigr)^{n-1} f(x),
\end{equation}
\begin{equation}
\hat \delta_N(x,k) \equiv \frac{-1}{(2 i k)^N} (\mathcal{A} - 2 i k \mathcal{B})^{-1} 
\mathcal{A} \,\hat c_N.
\end{equation}
[We used $\mathcal{B}_{(2)}^{-1} \, f(x) = f(x)$ in (9.9).]
From (9.7), we have the condition
\begin{equation}
\lim_{z \to -\infty} \mathcal{U}(x,z)\, \hat c_n(z) = 0.
\end{equation}
Expression (9.8) is valid if 
\begin{flalign}
\qquad \quad &\text{(i) \ the right-hand side of (9.9) makes sense for $n=1,2,\ldots,N$ \ \  and} &
\nonumber \\*
\qquad \quad &\text{(ii) \ condition (9.11) holds for $n=1,2,\ldots,N$.} &
\end{flalign}

From (6.9) and (9.2), we have
\begin{equation}
\mathcal{B}_{(2)}^{-1} \mathcal{A}_{(2)}^{\phantom{-1}} g(x,\xi)
=\frac{1}{\xi} \int_0^\xi \frac{\partial}{\partial x} g(x,\xi) \,d \xi
+ \frac{f(x)}{\xi}
\left[(1 - \xi^2) g(x,\xi) -g(x,0) \right].
\end{equation}
Using (9.13), we can calculate (9.9) as\cite{note4}
\begin{align}
&\hat c_1 = -f \mu^2, \qquad 
\hat c_2 = (- f' + f^2 \xi) \mu^2, \qquad
\hat c_3 = (- f'' + f^3 + 2 f f' \xi - f^3 \xi^2 ) \mu^2,
\nonumber \\*
&\hat c_4  
= [- f''' + 5 f^2 f' + (f'^2 + 2 f f'' - 2 f^4 ) \xi - 3 f^2 f' \xi^2 + f^4 \xi^3] \mu^2,
\qquad \hbox{etc.}
\end{align}
(The primes denote derivatives with respect to $x$.) 
In general, $\hat c_n$ has the form
\begin{equation}
\hat c_n
= \mu^2
\sum_{j=0}^{n-1} F_j \,\xi^j,
\end{equation}
where each $F_j$ is a polynomial in $f, f',f'', \ldots, f^{(n-1)}$.
It is obvious that $\hat c_n$ makes sense if $f \in C^{n -1}$. Therefore, condition (i) of (9.12) is satisfied if $f \in C^{N-1}$.

We need to examine condition (ii) of (9.12). Equations (9.15) and (3.21) give
\begin{equation}
\mathcal{U}(x,z)\, \hat c_n(z)
= \bigl[\hat T(x,z) \bigr]^2
\sum_{j=0}^{n-1} F_j(z)\bigl[\hat L(x,z)\bigr]^j.
\end{equation}
By using (9.16) and (C1)--(C4) of Appendix~C, we can easily check whether (9.11) holds or not.
When $f(-\infty)$ exists and is finite, 
(9.11) holds for $\mathrm{Im}\, k > 0$ if
\begin{equation}
\lim_{x \to -\infty} e^{\alpha x} f^{(m)}(x) = 0 
\qquad \text{for any \ $\alpha>0$ \ and for any \  $0\leq m \leq n-1$}.
\end{equation}
(Here, $f^{(m)}$ denotes the $m$\,th derivative of $f$.)
Equation (9.11) holds for $\mathrm{Im}\,k \geq 0$ if (9.17) is satisfied for $\alpha=0$ instead of $\alpha>0$.
When $f(-\infty) = \pm \infty$, (9.11) holds for $\mathrm{Im}\,k \geq 0$ if
\begin{equation}
\lim_{x \to -\infty}  e^{- \alpha \vert V(x) \vert}f^{(m)}(x) =0
\qquad \text{for any \ $\alpha>0$ \ and for any \  $0\leq m \leq n-1$}.
\end{equation}
When the potential is asymptotically periodic as in Sec.~VIII, Eq.~(9.11) holds for $\mathrm{Im}\, k > 0$ if (9.17) is satisfied.
This can be seen by using (9.16), (8.27), and (8.28).
 
The two conditions of (9.12) are satisfied for $\mathrm{Im}\,k>0$ if $f \in C^{N-1}$ and if either (9.17) or (9.18) holds with $n=N$. Then (9.8) is valid for $\mathrm{Im}\,k \geq 0$. [For $\mathrm{Im}\,k=0$, remainder term (9.10) should be interpreted as $\lim_{\epsilon \downarrow 0} \hat \delta(x,k+ i \epsilon)$.] 

For (9.8) to be an asymptotic expansion, it is also necessary that 
\begin{equation}
\lim_{\vert k \vert\to \infty} k^N \hat \delta_N(x,k) =0.
\end{equation}
Using (4.14), remainder term (9.10) can be expressed as
\begin{equation}
\hat \delta_N(x,k)
= \frac{-1}{(2 i k)^N} 
\int_{-\infty}^x dz\,\bigl[\hat T(x,z;k) \bigr]^2
\sum_{j=0}^{N} G_j(z)\bigl[\hat L(x,z;k)\bigr]^j,
\end{equation}
where each $G_j$ is a polynomial in $f, f', f'' \ldots, f^{(N)}$.
For fixed $x$ and $z$, we have 
\begin{equation}
\hat T(x,z;k) \sim \mu\, e^{i k (x-z)}, \qquad
\hat L(x,z;k) \sim \xi^2 \,e^{2i k (x-z)} \qquad \mbox{as  $\vert k \vert \to \infty$}.
\end{equation} 
Using (9.21), and also using the dominated convergence theorem and the Riemann-Lebesgue lemma, we can show that (9.19) holds in the half plane $\mathrm{Im}\,k \geq \epsilon$ with arbitrary $\epsilon>0$ if\cite{note5}
\begin{equation}
\int_{-\infty}^x e^{\alpha z}\,\vert f^{(m)}(z) \vert \,dz < \infty 
\qquad \text{for any \ $\alpha>0$ \ and for any \  $0\leq m \leq N-1$}.
\end{equation}
In particular, (9.19) holds in the half plane $\mathrm{Im}\,k \geq 0$ including the real axis if (9.22) holds for $\alpha=0$ instead of $\alpha>0$.
Condition (9.22) is not satisfied if $f(x)$ diverges exponentially or faster as $x \to -\infty$. In such cases, (9.19) still holds in the sector  $\epsilon \leq \arg k\leq \pi -\epsilon$ with $\epsilon>0$ if the derivatives of $f(x)$ are sufficiently well behaved as $x \to -\infty$.

\section{Concluding remarks}
In this paper, we studied low-energy expansion and high-energy expansion of the reflection coefficient $R_r(x,-\infty)$ by considering its generalized form $\hat R(x,-\infty;\mu,\xi)$.
The introduction of the two additional variables $\mu$ and $\xi$ gives a natural description of transmission and reflection processes and enables us to see clearly the meaning of the formulas used for the expansions.
In particular, the most important formula is Eq.(5.11) [or, equivalently, (4.14)]. 
This basic formula can be derived quite naturally in our generalized framework. 

The low- and high-energy expansions of $R_r(x,-\infty)$ are obtained from (7.10) and (9.8) by setting $\mu =1$ and $\xi=0$ [or $a=1$ and $W = V(x)$]. 
These asymptotic expansions are justified if 
 (i) each coefficient of the expansion makes sense, 
(ii) each coefficient of the expansion belongs to the appropriate domain of the operator $\mathcal{A} - 2 i k \mathcal{B}$, and (iii) the remainder term is of higher order in the expansion.
We have rigorously examined these conditions. Condition~(ii), especially, is essential to the validity of the expansions. (Conditions~(i) and~(iii) have been discussed in the previous works.\cite{analysis, high, low, asymptotic})  For the low-energy expansion, we have found that in all cases under consideration, condition~(ii) holds if condition~(i) is satisfied. For the high-energy expansion, condition~(ii) is not guaranteed by condition~(i).

The formalism introduced in this paper can be extended to a more general description of one-dimensional problems.
The operator $\mathcal{U}$ defined by (3.11) belongs to an infinite-dimensional representation of the Lie group $SL(2,\mathbf{C})$, with a representation space consisting of functions of $\mu$ and $\xi$. We can extend this representation to $SL(3,\mathbf{C})$. By utilizing this $SL(3,\mathbf{C})$ structure, it is possible to derive the expansions of the Green function directly, without using the reflection coefficients. 
This method will be exploited elsewhere.

\appendix
\section{DERIVATION OF (3.8)}
\renewcommand\theequation{A\arabic{equation}}
Let $\epsilon \equiv x - y$ be small. To first order in $\epsilon$, the diagrams contributing to $\hat T(x,y;\xi,\mu)$ are the ones shown in Fig.~4(a). 
In the second diagram, the integration over the position of $f$ is implied.
The contribution from these two diagrams is
\begin{equation}
\mu \exp(i k\epsilon) + \mu \xi\int_y^x f(z)\exp[i k \epsilon + 2 i k (x-z)] \, d z
= \mu + i k \mu \epsilon +\mu \xi f(x)\epsilon + o(\epsilon).
\end{equation}
Thus, $\hat T(x,y;\xi,\mu)=\mu + [i k \mu +\mu \xi f(x)]\epsilon + o(\epsilon)$.
Since $\hat T(x,x;\xi,\mu)=\mu$, we obtain
\begin{equation}
\left.\frac{\partial}{\partial x} \hat T(x,y;\xi,\mu) \right\vert_{y=x}
=\lim_{\epsilon \to 0} \frac{\hat T(x,y;\xi,\mu) - \mu}{\epsilon}
=i k \mu + \mu \xi f(x).
\end{equation}
This is the first equation of (3.8). 
Similarly, the remaining two equations are obtained from
\begin{equation}
\hat L(x,y;\xi)=\xi + [2 i k \xi + (\xi^2 - 1) f(x)]\epsilon + o(\epsilon),
\qquad
\hat R(x,y;\xi,\mu)=\mu^2 f(x) \epsilon +o(\epsilon),
\end{equation}
which can be seen from Fig.~4(b) and Fig.~4(c).

\section{DERIVATION OF (5.2)}
\renewcommand\theequation{B\arabic{equation}}
Let us define $\xi_\theta \equiv 
(\xi \cos \frac{\theta}{2} - \sin \frac{\theta}{2})/(\cos \frac{\theta}{2} + \xi \sin \frac{\theta}{2})$ and $\mu_\theta \equiv 
\mu/(\cos \frac{\theta}{2} + \xi \sin \frac{\theta}{2})$.
It is easy to show that
\begin{equation}
\frac{\partial}{\partial \theta}\xi_\theta
=-\frac{1 + \xi^2}{2}\frac{\partial}{\partial \xi} \xi_\theta=i \mathcal{J}_2 \xi_\theta, 
\qquad\frac{\partial}{\partial \theta}\mu_\theta
=-\frac{1 + \xi^2}{2}\frac{\partial}{\partial \xi} \mu_\theta
-\frac{\xi \mu}{2}  \frac{\partial}{\partial \mu} \mu_\theta =i \mathcal{J}_2 \mu_\theta.
\end{equation}
Since $\xi_{\theta=0}=\xi$ and $\mu_{\theta=0}=\mu$, from (B1) we have
$\exp(i \theta \mathcal{J}_2) \xi = \xi_\theta$ and
$\exp(i \theta \mathcal{J}_2) \mu= \mu_\theta$.
Setting $\theta=-\pi/2$ yields $ \exp(\frac{- i \pi}{2} \mathcal{J}_2)\xi=(1+\xi)/(1-\xi)$ and $\exp(\frac{- i \pi}{2} \mathcal{J}_2) \mu=\sqrt{2}\mu/(1-\xi)$.

\section{ASYMPTOTIC BEHAVIOR OF $\boldsymbol{\hat T(x,z)}$ AND $\boldsymbol{\hat L(x,z)}$ AS $\boldsymbol{z \to -\infty}$}
\renewcommand\theequation{C\arabic{equation}}
The behavior of $\hat T(x,z)$ and $\hat L(x,z)$ as $z \to -\infty$  differs according to the behavior of $f(x)$ as $x \to -\infty$.
As in Sec.~VII, we assume that $f(-\infty)$ and $f'(-\infty)$ exist.
We also assume that $\lim_{x \to -\infty} f'(x)/f(x)$ exists. (This limit may be infinite.)
We have the following asymptotic expressions of $\hat T$ and $\hat L$ as $z \to -\infty$. 
(For details, see Appendix~F of Ref.~\onlinecite{low}.)

\medskip

\noindent
If $f(-\infty)=+\infty$,
\begin{equation}
\hat T(x,z)=a \, C_1 \exp \! \left[-\frac{1}{2} V(z) + \eta_1(z)\right][1 + o(1)],
\qquad
\hat L(x,z)= -1 + o(1).
\end{equation}
If $f(-\infty)=- \infty$,
\begin{equation}
\hat T(x,z)=a \, C_2 \exp \! \left[\frac{1}{2} V(z) + \eta_2(z)\right][1 + o(1)],
\qquad
\hat L(x,z)= 1 + o(1).
\end{equation}
If $f(-\infty)=c \neq 0$,
\begin{equation}
\hat T(x,z) = a \, C_3 \exp \! \left[ \sqrt{c^2 - k^2} \,z + i \eta_3(z)\right][1 + o(1)],
\qquad
\hat L(x,z)= - \frac{i k + \sqrt{c^2 - k^2}}{c} + o(1).
\end{equation}
If $f(-\infty)=0$,
\begin{equation}
\hat T(x,z) = a \, C_4 \exp \! \left[ - i k z + i \eta_4(z)\right][1 + o(1)],
\qquad
\hat L(x,z)= C_5 \exp \! \left[ - 2 i k z + 2 i \eta_4(z)\right] + o(1).
\end{equation}
In the above expressions, each $C_i$ $(i=1,2,3,4,5)$ is a  function of $x$, $W$, and $k$ which does not depend on $z$, and each $\eta_i$ $(i=1,2,3,4)$ is a function of $z$ and $k$ such that
\begin{equation} 
\eta_i(z) = o(\vert z \vert) \qquad \mbox {as $z \to -\infty$}.
\end{equation}
When $\mathrm{Im}\,k=0$, the function $\eta_4(z)$ in (C4) is real valued, and $\vert C_5 \vert <1$. So it follows from (C4) that if $f(-\infty)=0$,
\begin{align}
&\lim_{z \to -\infty} \vert \hat T(x,z) \vert =0 
\quad \mbox{for $\mathrm{Im}\,k >0$},
\qquad
\lim_{z \to -\infty} \vert \hat T(x,z) \vert = \vert a C_4 \vert < \infty 
\quad \mbox{for $\mathrm{Im}\,k =0$},
\nonumber \\*
&\lim_{z \to -\infty} \vert \hat L(x,z) \vert =0 
\quad \mbox{for $\mathrm{Im}\,k >0$},
\qquad
\lim_{z \to -\infty} \vert \hat L(x,z) \vert = \vert C_5 \vert <1 
\quad \mbox{for $\mathrm{Im}\,k =0$}.
\end{align}


%



\begin{thebibliography}{33}
\bibitem{note1}
When $V(x)$ has a finite jump, the corresponding $V_\mathrm{S}(x)$ does not make sense. For the interpretation of such cases, see footnote~1 of Ref.~\onlinecite{low}.
\bibitem{note2}
In deriving equations~(2.5) and (2.12), we used $\det U=\alpha(k) \alpha(-k)-\beta(k) \beta(-k)=1$. This relation comes from the fact that the matrix in front of $U$ in Eq.~(1.5) is traceless.
\bibitem{note3}
References~\onlinecite{analysis, low, periodic, asymptotic} deal with $\bar R_r$ instead of $\hat R$, and the expansion of $\bar R_r(x,-\infty)$ is defined as
$\bar R_r = \sum_{n=0}^N (i k)^n \bar r_n + \bar \rho_N$. 
As can be seen from (5.16), the relation between $\hat r_n, \hat \rho_N$ and $\bar r_n, \bar \rho_N$ is given by $\hat r_0 = a^2 \bar r_0 + a^2 \xi$, $\hat r_n = a^2 \bar r_n$ for $n \geq 1$, and $\hat \rho_N = a^2 \bar\rho_N$.
\bibitem{note4}
In Refs.~\onlinecite{analysis, high}, the expansion of $\bar R_r$ is defined as
$\bar R_r = -\xi + \sum_{n=1}^N \bar c_n/(2 i k)^n + \bar \delta_N$.
The relation between $\hat c_n, \hat \delta_N$ and $\bar c_n, \bar \delta_N$ is given by $\hat c_n/\mu^2 = \bar c_n/(1-\xi^2)$ and $\hat \delta_N/\mu^2 = \bar \delta_N/(1 - \xi^2)$.
\bibitem{note5}
Note that condition (9.22) involves $f^{(N-1)}$ but not $f^{(N)}$. 
If (9.22) holds, it can be shown that $\lim_{\vert k \vert \to \infty}(2 i k)^{N} \hat \delta_{N-1} = \hat c_N$. Since $\hat \delta_{N-1}=(2ik)^{-N} \hat c_N + \hat \delta_N$, this implies (9.19).
\end{thebibliography}
\end{document}